\documentclass[prl,twocolumn]{revtex4-1}
\usepackage{graphicx}
\usepackage{subfigure}
\usepackage{booktabs}
\usepackage{array}
\usepackage{amsmath}
\usepackage{amssymb}
\usepackage{wasysym}
\usepackage{bm}
\usepackage{placeins}
\usepackage[dvipsnames]{xcolor}

\usepackage{hyperref}
\hypersetup{
    colorlinks=true,       
    linkcolor=blue,          
    citecolor=blue,        
    filecolor=magenta,      
    urlcolor=cyan           
}

\bibpunct{[}{]}{,}{n}{}{}
\newcolumntype{C}{>{$}c<{$}}
\AtBeginDocument{
\heavyrulewidth=.08em
\lightrulewidth=.05em
\cmidrulewidth=.03em
\belowrulesep=.65ex
\belowbottomsep=0pt
\aboverulesep=.4ex
\abovetopsep=0pt
\cmidrulesep=\doublerulesep
\cmidrulekern=.5em
\defaultaddspace=.5em
}

\newcommand{\ra}[1]{\renewcommand{\arraystretch}{#1}}

\def\k{\ensuremath \bm{k}}

\def\v{\ensuremath \bm{v}}
\def\tp{\ensuremath t^{\prime}}
\def\sn{\ensuremath \text{sn}}
\def\cn{\ensuremath \text{cn}}
\def\dn{\ensuremath \text{dn}}
\def\sd{\ensuremath \text{sd}}

\def\sc{\ensuremath \text{sc}}

\def\sech{\ensuremath \text{sech}}
\def\vxt{\ensuremath \tilde{v}_x}

\def\Kp{\ensuremath K^{\prime}}

\def\up{\ensuremath u^{\prime}}
\def\O{\ensuremath \mathcal{O}}

\begin{document}

\title{The Hall number across a van Hove singularity}
\author{Akash V. Maharaj${}^{1}$,  Ilya Esterlis${}^{1}$, Yi Zhang${}^{1,2}$, B.J. Ramshaw${}^{3, 4}$  and S. A. Kivelson${}^{1}$ }
\affiliation{${}^{1}$Department of Physics, Stanford University, Stanford, California 94305, USA.}
\affiliation{${}^{2}$Department of Physics, Cornell University, Ithaca, NY 14853, USA}
\affiliation{${}^{3}$Los Alamos National Laboratory, Los Alamos, New Mexico 87545, USA.}
\affiliation{${}^{4}$Laboratory of Atomic and Solid State Physics, Cornell University, Ithaca, NY 14853, USA}
\date{\today}

\begin{abstract}
In the context of the relaxation time approximation to Boltzmann transport theory, we examine the behavior of the Hall number, $n_H$, of a metal in the neighborhood of a Lifshitz transition from a closed Fermi surface to open sheets. We find a universal non-analytic dependence of $n_H$ on the electron density in the high field limit, but a non-singular dependence at low fields.  
The existence of an assumed nematic transition produces a doping dependent $n_H$ similar to that observed in recent experiments in the high temperature superconductor YBa$_2$Cu$_3$O$_{7-x}$.
\end{abstract}

\maketitle

\textbf{\textit{Introduction.}}---In the absence of superconductivity or exotic fractionalized phases, the low energy elementary excitations of a conducting system are typically the well-known quasiparticles of Fermi liquid theory.  In sufficiently clean systems, much about the character of these excitations, and in particular, information concerning the geometry and topology of the Fermi surface, can be inferred most sensitively from transport experiments. Specifically, in many circumstances, the Hall number, $n_H\equiv (B/e)(1/\rho_{xy})$, in the $T\to 0$ limit can give information about the volume (area in $2D$) enclosed by the Fermi surface. 
  \cite{Shoenberg:1984,abrikosov1988}. From this, one may extract insight concerning the existence of a putative broken symmetry state that ``reconstructs'' the Fermi surface. For example, density wave order that breaks translational symmetry, changes not only the topology of the Fermi surface, but the volume enclosed as well. In contrast, the constraints of Luttinger's theorem seemingly imply that Fermi surface changes produced by translation symmetry preserving orders, such as Ising nematic order, will be invisible to a measurement of the Hall number.

There are, however, important caveats to using the Hall number as a proxy for the electron density of a metal. In the absence of Galilean invariance, it is only the $B \rightarrow \infty$ limit of the Hall number that corresponds to the carrier density\cite{abrikosov1988}. The $B \rightarrow 0$ limit of the Hall number is sensitive to the momentum dependence of the Fermi velocity, and is related in a complicated way \cite{ong1991geometric} to the dominant scattering processes and curvature of the Fermi surface. For open Fermi surfaces, the Hall number is in general a non-universal quantity, and is not related to the density in any simple fashion in either the strong or weak field limit. In fact, little is known about the critical behavior of the Hall number at the topological Lifshitz phase transition between open and closed Fermi surfaces. While there is intuitively no reason to expect singular behavior in the limit $B \rightarrow 0$, since the Fermi surface is locally unchanged across the van Hove singularity, there is every reason to expect singular behavior at high fields, where quasiparticles exhibit many cyclotron orbits before being scattered, and so are sensitive to the global topology of the Fermi surface. 

In this Letter, we address these issues via exact solution of the Boltzmann equation in the relaxation time approximation for a two dimensional nearest-neighbor tight binding model, and by numerical solution of models with other band-structures.  We report results in the $T\to 0$ limit under the assumption that the semiclassical approximation is valid, {\it i.e.} $\omega_c/\epsilon \ll 1$ where $\omega_c \propto B$ is the cyclotron energy and $\epsilon$ is the smallest significant energy scale characterizing the band-structure at energies near the chemical potential, $\mu$. Subject to this constraint, we will discuss our results in the high and low field limits, $\omega_c\tau \gg 1$  and $\omega_c\tau \ll 1$ respectively, where $\tau$ is the relaxation time. In the high field limit $n_H$ is non-analytic at the point of transition from a closed to an open Fermi surface. Specifically, $n_H =n$ in a metal with only closed Fermi pockets, while for open Fermi sheets $n_H$ is not simply related to $n$; we find that it exhibits a non-analytic evolution, 
\begin{align}
n_{H} - n \propto \frac{n_c}{\log{|n - n_c|}},
 \label{nHofn}
\end{align}
upon approach to the Lifshitz transition at  $n= n_c$. \footnote{To be precise, in the presence of multiple Fermi surfaces, note that $n_H = -(n_{e} - n_{h})$, where $n_e$ is the area enclosed by electron pockets, and $n_h$ is the area enclosed by hole pockets}
Conversely, at low fields, $n_H$ is smooth as a function of density in the neighborhood of $n_c$.

Suggestively similar behavior of $n_H$ has recently been reported\cite{badoux2016change} in the hole doped cuprate superconductor YBa$_2$Cu$_3$O$_{7-x}$ (YBCO). There, $n_H$ was found to rise sharply on approach to a critical hole doping of $p=p^{\star} \approx 20\%$, although the very high values of $H_{c2}$ have  precluded measurements below approximately 40K.  A somewhat similar sharp increase of $n_H$ as $p$ approaches a critical value near optimal doping was reported previously in Bi-2201\cite{ando1997normal, balakirev2003signature} and LSCO\cite{balakirev2002underdoped, balakirev2009quantum}, where the lower critical fields permitted experiments at much lower temperatures.  In these latter studies, the Hall number decreases at higher doping (i.e. $n_H$ is peaked at $p^\star$), while more recent studies of LSCO and LNSCO\cite{daou2009linear, collignon2016fermi} have inferred that $n_H$ saturates at a value $n_H\sim (1+p)$ for $p>p^\star$.  (These observations are yet to be reconciled.)
 
The idea that measurements of $n_H$ performed in high enough fields to quench superconductivity could be used to identify a quantum critical point (QCP) was introduced by Chakravarty {\it et al.} \cite{chakravarty2002sharp} in the context of a $d$-density-wave (dDW)  QCP, and soon after by Kee {\it et al.} \cite{kee2003signatures} for a model of a metal undergoing a first order nematic-to-isotropic transition. In both cases, the Hall number was found to decrease significantly in the ordered phase.  Here, we show that a singular drop in $n_H$ is also consistent with a continuous nematic phase transition. This result may be applicable to YBCO assuming that its low temperature in-field properties can be treated in the context of Fermi liquid theory\footnote{The assumption that Fermi liquid theory is valid in the cuprates at low $T$ and moderate $B$ fields is supported by Quantum Oscillations measurements, where Lifshitz-Kosevich-like temperature dependence has been reported.\cite{sebastian2010fermiliquid}}.


\textit{\textbf{Chambers' Formula.}}---
We compute the magnetotransport using Chambers' expression for the conductivity tensor\cite{shockley1950effect,chambers1952kinetic}. This is a formally exact integral solution to the Boltzmann equation in the relaxation time approximation, correct to all orders in $B$. The conductivity tensor at zero temperature in $d$ dimensions takes the form ($\hbar = 1$)
\begin{align}
\sigma_{\alpha\beta} &= \frac{e^{2}}{(2\pi)^d} \int \frac{dS}{|\bm{v}|} v_{\alpha}(0)  \int^{0}_{-\infty}d\tp  v_{\beta}(\tp) e^{\tp/\tau}
\end{align}
where $\tau$ is the scattering time, the first integral is over the Fermi surface (FS), and the effect of the magnetic field is included implicitly via the quasiparticle velocities $\bm{v}(t)$ along a cyclotron orbit. To evaluate this expression requires that for each point $\k$ on the FS, we calculate $\bm{v}(t) = \bm{\nabla}_{\k} \varepsilon(\k(t))$, where $\k(t)$ evolves according to the Lorentz force law: $\dot{\k} = -e \v \times \bm{B}$ \cite{Goddard:2004}. The solutions are generically periodic with period $T$, and therefore in $d=2$,
\begin{align}
\sigma_{\alpha\beta}&= \frac{e^{3}B}{(2\pi)^2 }\int^{T}_{0} dt\,\, v_{\alpha}(t) \int^{t}_{-\infty} dt^{\prime}\,\,\, v_{\beta}(t^{\prime})e^{(t^{\prime} - t)/\tau}.
\label{chambers}
\end{align}
  

\textit{\textbf{Nearest-neighbor tight-binding model}}.---
We consider spinless electrons on a square lattice,  
\begin{equation}
H = \sum_{\k} \varepsilon(\k) c^{\dag}_{\k}c_{\k},
\end{equation}
where $H$ is the Hamiltonian, $c^{\dag}_{\bm{k}}$ creates an electron with Bloch wave-number $\k$,  $t_x$ and $t_y$ are the hopping strengths on  $\hat{\bm{x}}$ and $\hat{\bm{y}}$ directed bonds, and 
\begin{align}
\varepsilon(\k) &= -2t_x \cos{k_x} - 2t_y\cos{k_y}.
\end{align}

Chambers' formula for this model can be evaluated exactly \footnote{This form of exact solution is similar to that in Schofield et al.\cite{schofield2000quasilinear}, where weakly coupled 1d chains were considered. There, Galilean invariance is present in one direction so the Hall number is always equal to the density.}. The solutions for the quasiparticle velocities at a given chemical potential $\mu$ are rational fractions of Jacobian elliptic functions, with the corresponding cyclotron frequency given by:
\begin{align}
\omega_c = \begin{cases}
\frac{\pi}{2K(\kappa)}\omega_0 &\,\text{closed orbits, } |\mu| > \mu_c\\
\frac{\kappa\pi}{2K(1/\kappa)}\omega_0&\,\text{open trajectories, }|\mu| \le \mu_c
\end{cases}
\label{eq:trueOmegaC}
\end{align}
where $\omega_0 = eB\sqrt{4t_xt_y} $ is a `bare' cyclotron frequency, and $K(\kappa)$ is the complete elliptic integral of the first kind, with elliptic modulus given by 
\begin{align} 
\kappa = \sqrt{\frac{\mu^2_0 - \mu^2}{\mu^2_0 - \mu^2_c}}.
\end{align}
Here, $\mu_0 = 2(t_x + t_y)$ is half the bandwidth, and the van Hove singularities occur at  $\mu = \pm \mu_c = \pm 2(t_y - t_x)$. 

The integral in Eq. \ref{chambers} is tractable provided Fourier series expansions for the quasiparticle velocities can be computed. The gory details of the lengthy, but straightforward manipulations needed to achieve this are presented in the Supplemental Material. The final results for the conductivities of the $i = e, o, h$ (electron, open and hole pockets respectively) are expressible as rapidly convergent infinite series over Fourier coefficients of the quasiparticle velocities:
\begin{align}
\sigma^{i}_{xx} &= \frac{2\sigma_0
 }{K}   \sum_{m}  \frac{\sech^{2}{\left(\frac{m\pi K^{\prime}}{2K}\right)}\sin^2{\left(\frac{m\pi u_{i}}{2K}\right)}}{1 +\left(m\omega_c \tau \right)^2},\label{eq:sxx}\\
\sigma^{i}_{yy} &=  \frac{\sigma_0\delta_{i,o}}{K} +   \frac{2\sigma_0}{K}\sum_{m}  \frac{\sech^{2}{\left(\frac{m\pi K^{\prime}}{2K}\right)}\cos^2{\left(\frac{m\pi u_{i}}{2K}\right)}}{1 + \left(m\omega_c \tau \right)^2},\label{eq:syy}\\
\sigma^{i}_{xy} &=   \frac{\sigma_0}{K}  \sum_m  \, \,\frac{(m\omega_c\tau)\,\,\sech^{2}{\left(\frac{m\pi K^{\prime}}{2K}\right)}\sin{\left(\frac{m\pi u_{i}}{K}\right)}}{1 + \left(m\omega_c \tau \right)^2},\label{eq:sxy}
\end{align}
with $\sigma_0=e^2\tau\sqrt{4t_xt_y}$. For closed pockets ($i = e, h$), the sums are over positive odd integers, while for open Fermi surfaces ($i = o$), the sum is over positive even integers. We have used the shorthand notation $K \equiv K(\kappa)$ for closed pockets, and $K^{\prime} \equiv K(\sqrt{1-\kappa^2})$; for open surfaces we substitute $K(\kappa)\rightarrow \frac{1}{\kappa}K(1/\kappa)$ and likewise for $K^{\prime}$. Finally, the parameters $u_i$ are defined implicitly as
\begin{equation}
\begin{aligned}
\sn(u_{e/h},\kappa) &= \sqrt{(\mu_0 - \mu_c)/(\mu_0 \mp \mu)}\\
\sn(\kappa u_{o},1/\kappa) &= \sqrt{(\mu_0 + \mu)/(\mu_0 + \mu_c)}
\end{aligned}
\end{equation}
where $\sn(u,k)$ is a Jacobian elliptic function.
 
\begin{figure}
\includegraphics[width=0.4\textwidth]{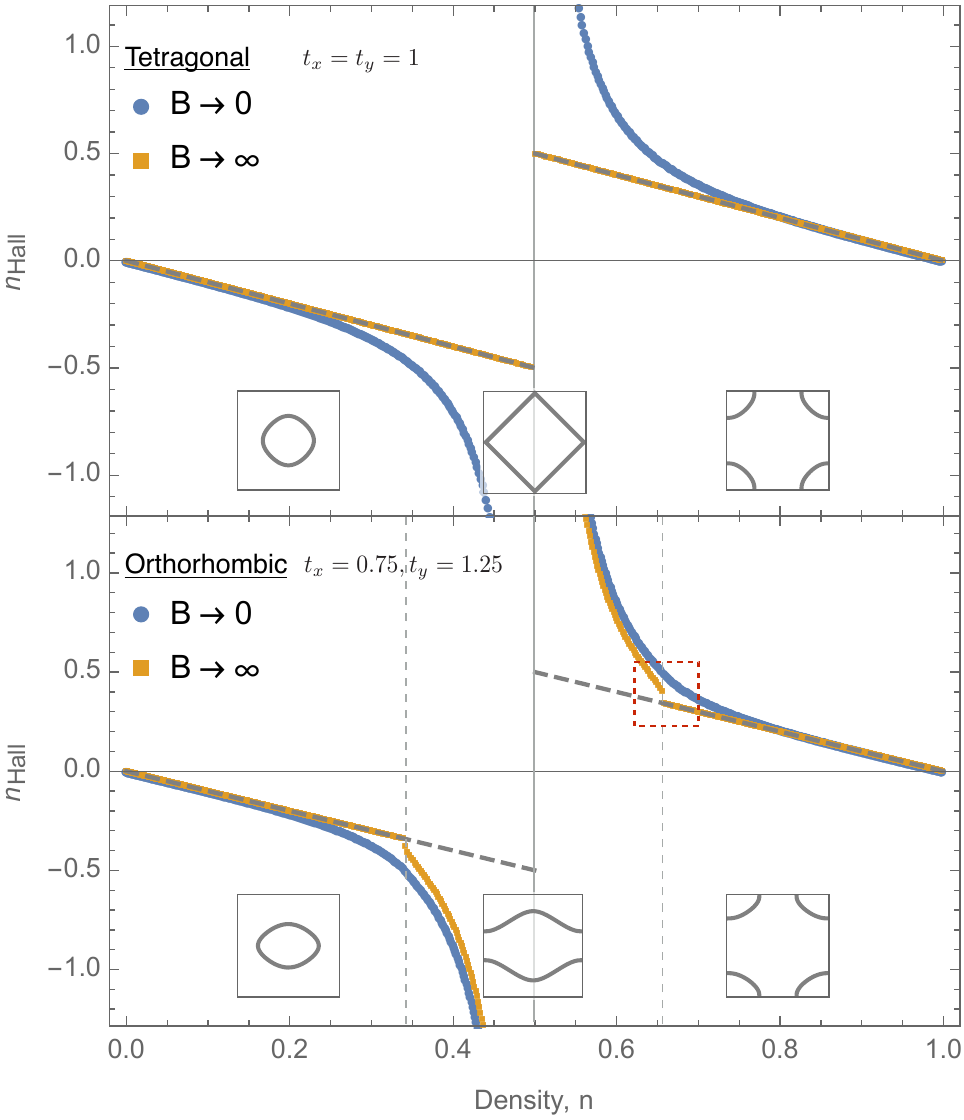}
\caption{The Hall number $n_H$ in the $B \rightarrow 0$ and $B \rightarrow \infty$ limits versus density for tetragonal and orthorhombic systems. Insets show the Fermi surfaces(FS) at densities $n = 0.2, 0.5,$ and $0.8$. When the FS is closed, $n_H$ is exactly equal to the carrier density  (dashed lines) in the large $B$ limit, while for $B\rightarrow 0$, $n_H$ deviates strongly from the dashed line near to the van Hove fillings where the topology of the FS changes. For orthorhombic systems, a sharp non-analyticity exists in the high field $n_H$ at the van Hove fillings. The red box shows the critical region examined in Fig.~\ref{fig:asymptotics}}  \label{fig:mudep_comparison}
\end{figure}

The Hall number is computed from the conductivity tensor as
\begin{align}
\frac{1}{n_H} = \frac{1}{eB}\left[\frac{\sigma_{xy}}{\sigma_{xx}\sigma_{yy} +\sigma^{2}_{xy}}\right].
\end{align}
Figure 1 shows both the $B\rightarrow 0$ and $B\rightarrow \infty$ limits of this expression, for a tetragonal ($t_x = t_y$) and orthorhombic ($t_x < t_y$) systems. For closed FS's in both tetragonal and orthorhombic systems, the high field $n_H$ (yellow points) corresponds to the density of electrons or holes. In the low field limit (blue points), $n_H$ is only equivalent to the carrier density near the band edges, where Galilean invariance is approximately recovered. For generic fillings, the low field Hall number is inequivalent to the electron density; it in fact diverges near the band center, where the FS curvature vanishes. For orthorhombic systems, when there is an open FS, $n_H$ is not equivalent to the density even in the high field limit. There is a sharp non-analyticity at the Lifshitz transition in the high field limit, but not in the low. Note that $n_H$ diverges (\textit{i.e.} the Hall coefficient and hence the Hall voltage vanishes) at the point of particle-hole symmetry, $n = 0.5$, even though the evolution of the open FS is in no way singular at this point\cite{dagan2016fermi}.


\textbf{\textit{Critical behavior.}}---
So long as there are no open pieces of Fermi surface, the Hall number in the infinite field limit is equal to the (net) area enclosed by the Fermi surface(s). However for open surfaces, it follows from expressions for the magneto-conductivity (Eq.~\ref{eq:sxx} - \ref{eq:sxy}) that 
 $n_H \to n^{(o)}_{H} $ where
\begin{align}
n^{(o)}_{H} &= -2\frac{S(u_o)}{K(1/\kappa)S'(u_o)} - K(1/\kappa)S'(u_0),
\label{exact}
\end{align}
and
\begin{align}
S(u_o) &\equiv \frac{1}{\pi^2} \sum^{\infty}_{m=1} \frac{1}{m^2}  \sech^{2}{\left(\frac{m\pi K^{\prime}(1/\kappa)}{K(1/\kappa)}\right)}\sin^{2}{\left(\frac{m\pi u_{o}}{K(1/\kappa)}\right)}.
\nonumber
\end{align}

\begin{figure}[t]
\includegraphics[width=0.48\textwidth]{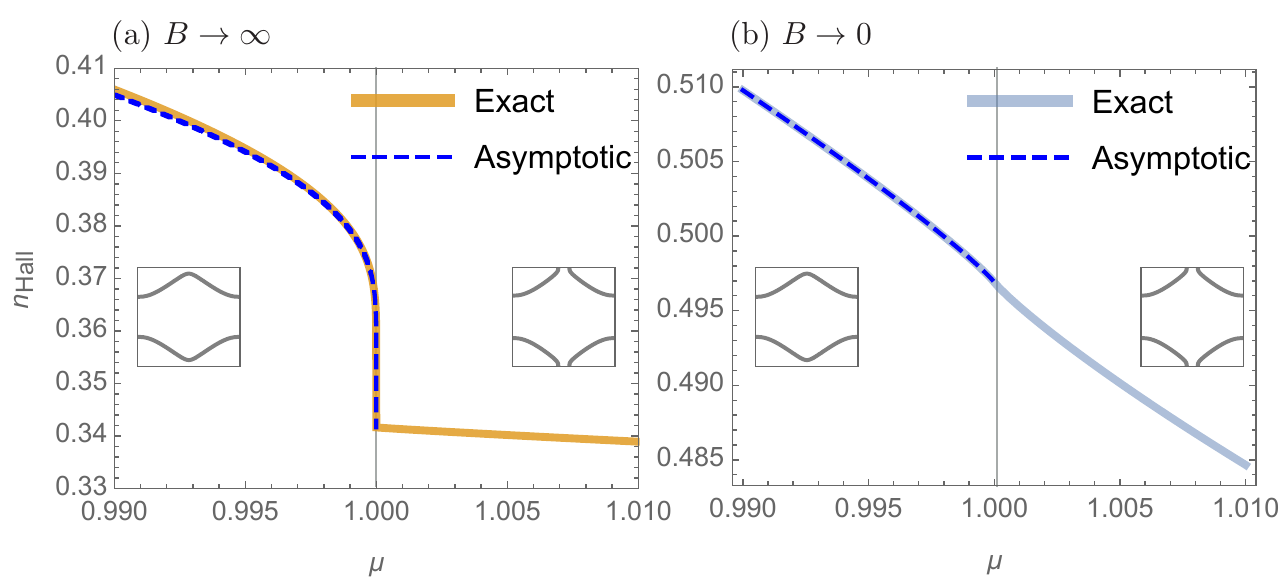}
\caption{Asymptotic behavior of the high and low field Hall number across the van Hove singularity at $\mu_c  = +2(t_y - t_x) = 1$ (i.e. $t_y = 1+\phi$ and $t_x = 1-\phi$, with $\phi = 0.25$). For $\mu = \mu_c -\delta\mu$, the sharp non-analyticity in the high field limit is of the form in Eq~\ref{eq:HallAnomaly}, while in the low field limit it is weaker, and of the form $\delta\mu \log{|\mu_c/\delta\mu|}$. Insets are schematics of the FS on either side of $\mu_c$.}  \label{fig:asymptotics}
\end{figure}

The particle-hole symmetry of the present model relates the behavior at density $n$ to that at density $1-n$, so without loss of generality we focus on the more-than-half-filled band, $1/2<n<1$. Near the van Hove, where $\mu =( \mu_c -\delta \mu)$ with $0<\delta \mu \ll \mu_c$, the sum can be evaluated up to small corrections in powers of $\delta \mu/\mu_c$ with the result $n^{(o)}_{H}= n_c +\delta n_H$ where $n_c$ is the density at $\mu=\mu_c$, and
\begin{align}
\delta n_{H} (\mu) = \frac{n_c\, C_1}{\log{\left|C_2 \delta\mu/\mu_c \right|}} + \O\left(\frac{\delta\mu}{\mu_c}\right),\label{eq:HallAnomaly}
\end{align}
in which $C_{1}$ and $C_{2}$ are $\mu$-independent dimensionless constants with complicated dependences on $t_x/t_y$.  (Explicit expressions are given in the Supplemental Material.) A comparison between the exact $\mu$ dependence of $n_H^{(o)}$ from  Eq. \ref{exact} and the asymptotic expression in Eq. \ref{eq:HallAnomaly} is shown in Figure~\ref{fig:asymptotics}(a). 

It is illuminating to express $n_H$ as a function of the electron density, $n$.  In $2D$, the {\it density of states} diverges logarithmically at the van-Hove point, but the density is continuous, with a weakly non-analytic form 
\begin{align}
n(\mu) - n \propto  \delta \mu\ \log{|\delta\mu/\mu_c |}.\label{eq:densAnomaly}
\end{align}
Consequently, $n_H(n)$, given in Eq. \ref{nHofn}, behaves in much the same  way as $n_H(\mu)$.

In the low field limit, $n_H(\mu)$ is again expressible in terms of infinite series and the sums can be performed, as discussed in the Supplemental Material. While the resulting expression is still singular at $\mu_c$, the singularity is much weaker as shown in Figure ~\ref{fig:asymptotics}(b); it simply reflects the logarithmic divergence of the density of states. Consequently, both $n_H(n)$ and its first derivative are continuous at $n=n_c$. \footnote{
{Note that close enough to the Lifshitz transition, the condition $\omega_c < \epsilon$ necessarily breaks down. Moreover, the assumption of constant $\tau$ becomes questionable.}}

Concerning experimental realizations, one can tune across the Lifshitz transition either by changing the chemical potential $\mu$, or the orthorhombicity, $\phi$.  $\mu$ is tuned by changing the electron concentration, either by doping or possibly by gating. $\phi$ can be directly varied by application of appropriate strain\cite{hicks2014strong, burganov2016strain, steppke2016strong}, or indirectly in systems which spontaneously break $C_4$ symmetry, by perturbations that affect the magnitude of the nematicity. 
\begin{figure}[t]
\includegraphics[width=0.5\textwidth]{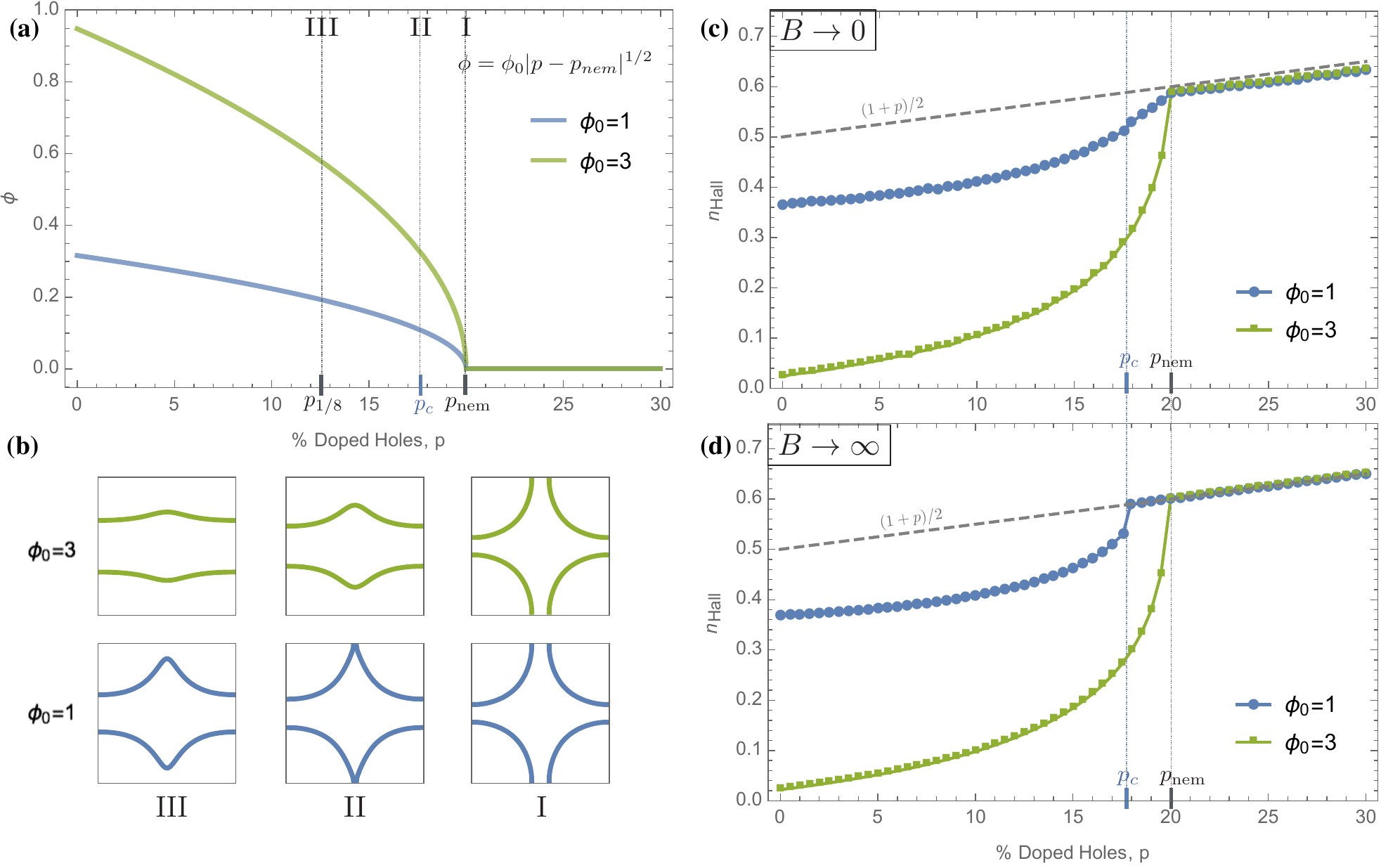}
\caption{The Hall number computed numerically, as a function of hole doping, $p$ for a model in which a nematic phase transition occurs at $p_{\text{nem}} = 20\%$.\textbf{(a)} The doping evolution of the nematic order parameter $\phi$. \textbf{(b)} Fermi surfaces as a function of doping: the Lifshitz transition is generically separated from $p_{nem}$; for $\phi_0 = 1$ it occurs at $p_c < p_{\text{nem}}$, however for the stronger onset ($\phi_0 =3$), it is too close to $p_{\text{nem}}$ to be resolved. \textbf{(c)} and \textbf{(d)} The sharpness of the drop in $n_H$ in both the strong and weak field limits is controlled by $\phi_0$. Longitudinal resistivities are shown in the Supplemental Material.}  \label{fig:nematicNumerics}
\end{figure}


\textbf{\textit{Possible relevance to the cuprates}}.--- 
The cuprate phase diagram is complex, with multiple ``intertwined'' orders. This complicates attempts to associate particular features of the transport, even apparent singularities, with specific ordering tendencies.  Given the considerable evidence of a tendency to nematic order in the cuprates\cite{andoPRL2002, hinkovScience2008, lawler2010intra, cyr2015two, ramshaw2017rotational}, we have undertaken to show that a nematic transition {\it could} produce a doping dependence of the Hall number similar to that seen in experiment.  However, this is merely a consistency check; similar behavior of $n_H$ was predicted on the basis of an assumed dDW transition\cite{chakravarty2002sharp}, and has been postdicted on the basis of assumed transitions involving spin or charge density wave (CDW) order\cite{storey2016hall, sachdev2016spin,harrison2016number}, spiral antiferromagnetism\cite{eberlein2016fermi}, or a transition to an  
``FL* phase''\cite{patel2016confinement,chatterjee2016fractionalized}. 

\begin{figure}[t!]
\includegraphics[width=0.45\textwidth]{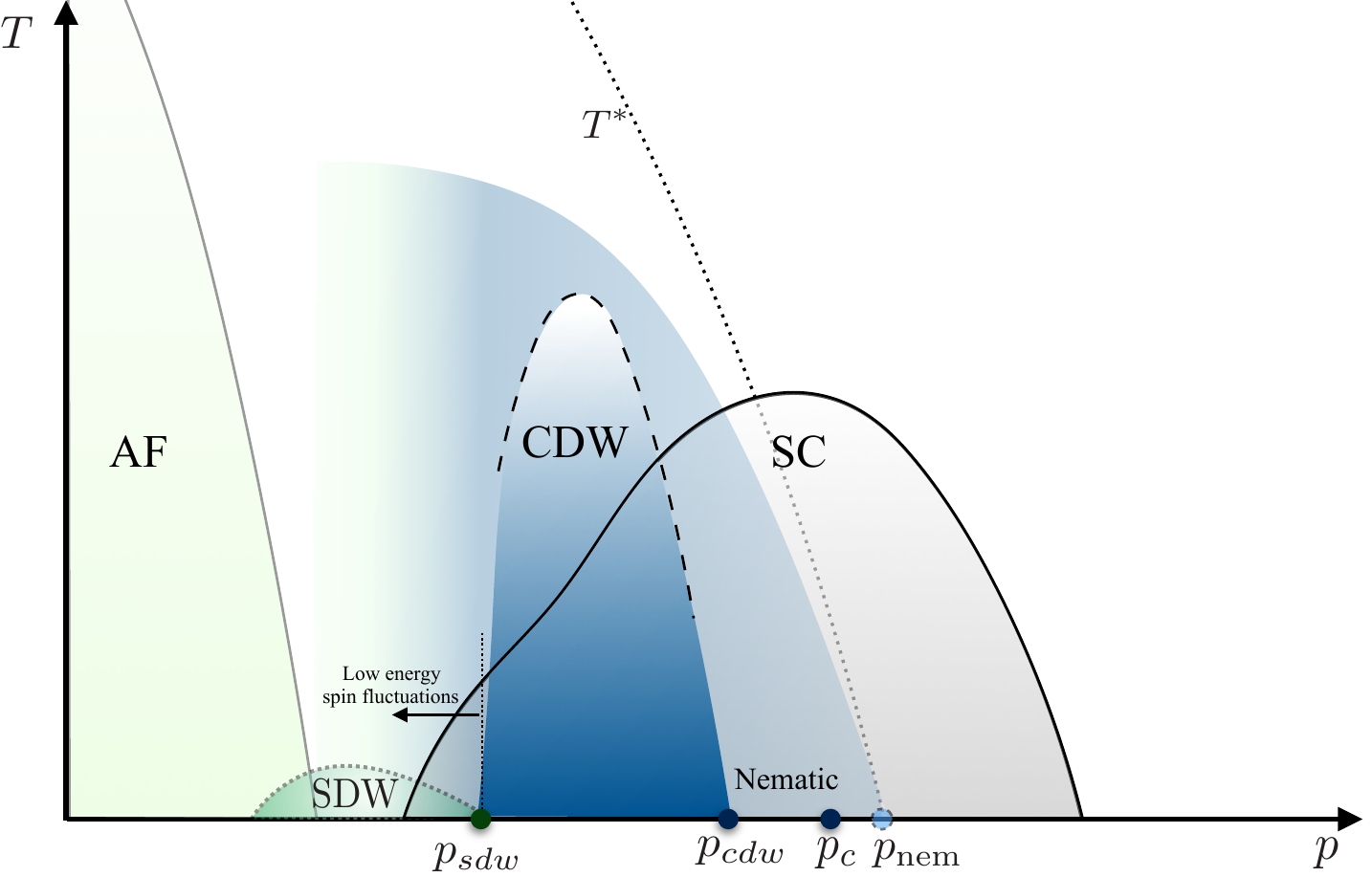}
\caption{A speculative zero field phase diagram of an ideal cuprate with a nematic phase included. In a tetragonal cuprate, $p_\text{nem}$ is a nematic quantum critical point, which in YBCO would be rounded by weak orthorhombicity. Here, we have considered $p_c$ to be a Lifshitz transition, which generically occurs inside the nematic phase. There is then a continuous transition to a unidirectional CDW. SDW and AF represent different forms of magnetic order, although in the presence of disorder, the SDW is typically manifest as a spin-glass.  All the ordered phases occur below a pseudogap crossover temperature, $T^*$.}  \label{fig:schematicpd}
\end{figure}

To capture something of the electronic structure of the cuprates, we have considered an electronic dispersion of the form $\varepsilon(\k) = -2t(1-\phi)\cos{k_x} - 2t(1+\phi)\cos{k_y} + 4t^{\prime}\cos{k_x}\cos{k_y}$, with $t^\prime = 0.4t$. Here $\phi$ is the nematic order parameter, which we assume has a mean-field-like dependence $\phi = \phi_0[ p_{\text{nem}}-p]^{1/2}$  on the doped hole concentration, $p$, with $p<p_{\text{nem}} \approx 20\%$. With second neighbor tunneling ($t^\prime$),  Chambers' formula becomes analytically intractable, so we obtain results numerically.\footnote{Details of the model and the method of solution are presented in the Supplemental Material.} 
 
As Fig.~\ref{fig:nematicNumerics} illustrates, the relation of the Hall number to the FS area differs at high and low fields. From the ratio of $\rho_{xy}$ to $\rho_{xx}$ taken from the Hall measurements of Badoux et al. \cite{badoux2016change} on YBCO at $p = 0.205$, we estimate $\omega_c \tau \approx 0.17$ for $B = 90 T$ and $T = 50 K$; from quantum oscillation measurements at $p=0.152$ \cite{Ramshaw:2015} we estimate $\omega_c \tau \approx 0.5$ for $B = 90 T$ and $T = 1.5 K$. Both estimates place the YBCO Hall measurements in the low-field limit. Indeed, the low-field curves in Fig.~\ref{fig:nematicNumerics} resemble the behavior measured in YBCO. 

To place these results in context, Fig. \ref{fig:schematicpd} shows a speculative phase diagram of an ideal cuprate. 
There is considerable evidence of the existence of a  QCP at $p^*=0.2$ associated with the termination of
a pseudogap crossover line $T^*$, as shown. Various ordering tendencies occur in the pseudogap regime.  While it seems likely that  CDW order terminates at  lower doping, $p_{cdw} < p^*$,  vestigial nematic order is more robust\cite{niePNAS2014, nie2017vestigial} - we have shown it terminating at $p_{nem} \approx p^*$. Moreover, recent work \cite{EunAhPRB2016,lederer2016superconductivity} has shown that nematicity can account for some  of the pseudogap phenomenology, including Fermi arcs and bad metal behavior.  A notable aspect of this proposal is the existence of a Lifshitz  transition at $p_c$, at which the Fermi surface topology changes; in contrast to a nematic transition, this  is sharply defined only at $T=0$.

There are several testable consequences of this scenario:  1) The presence of open Fermi surfaces results in large resistive anisotropies as well as non-saturating magnetoresistance in the `open' direction. 2) A continuous transition at $p_{cdw}<p_{\text{nem}}$ to a charge density wave (CDW) ordered phase is possible only if the CDW is unidirectional. 
3) The nematic transition is replaced by a crossover in an orthorhombic crystal, such as YBCO;  however, the Lifshitz transition remains a sharply defined QCP. An attractive aspect of this scenario is that optimal doping is proximate to both a Lifshitz and a nematic QCP, both of which have been shown to enhance $T_c$ under appropriate circumstances. 

\begin{acknowledgments}
We acknowledge insightful discussions with Gregory Boebinger, Sudip Chakravarty, Samuel Lederer, and Louis Taillefer. 
AVM, IE, YZ and SAK were supported in part by NSF grant \#DMR 1265593 at Stanford. YZ also acknowledges support through the Bethe Postdoctoral Fellowship. BJR acknowledges funding by the U.S. Department of Energy Office of Basic Energy Sciences ÒScience at 100 TÓ program.
\end{acknowledgments}

\bibliographystyle{apsrev4-1} 
\bibliography{refs}


\widetext
\clearpage
\begin{center}
\textbf{\large Supplemental Material: The Hall Number across a van Hove singularity}
\end{center}
\setcounter{equation}{0}
\setcounter{figure}{0}
\setcounter{table}{0}
\makeatletter
\renewcommand{\theequation}{S\arabic{equation}}
\renewcommand{\thefigure}{S\arabic{figure}}
\renewcommand{\bibnumfmt}[1]{[S#1]}
\renewcommand{\citenumfont}[1]{S#1}
\tableofcontents



\section{Numerical Solutions for a next-nearest neighbor tight binding model}
With the inclusion of second neighbor hopping on the square lattice, the Chambers formula is no longer analytically tractable. 
Our numerical solutions proceed by numerically solving for the time evolution of quasiparticles on the Fermi surface, and by discretization of the Chambers formula in Eq. 3 of the main text:

\begin{align}
\sigma_{\alpha\beta} &= \frac{e^{3}B}{(2\pi)^2 }\int^{T}_{0} dt\,\, v_{\alpha}(t) \int^{t}_{-\infty} dt^{\prime}\,\,\, v_{\beta}(t^{\prime})e^{(t^{\prime} - t)/\tau}.
\end{align}
We discretize these integrals using numerical solutions for the quasiparticle's velocities as a function of discrete time $n \Delta t$ where $N\Delta t = T$. The periodic nature of the quasiparticle orbits means that the second integral can be truncated to one period, with an additional infinite sum 
\begin{align}
\sigma_{\alpha\beta} &= \frac{e^{3}B}{(2\pi)^2 } \Delta T^2 \sum^{N}_{m} v_{\alpha}(m\Delta t) \sum^{N}_{n} v_{\beta}(n\Delta t)e^{(n-m)\Delta t/\tau}\left( 1 + e^{-T/\tau} + e^{-2T/\tau} + \ldots\right)\\
&=  \frac{e^{3}B}{(2\pi)^2 }\frac{ \Delta T^2}{1 -e^{-T/\tau}} \sum^{N}_{n, m} v_{\alpha}(m\Delta t) v_{\beta}(n\Delta t)e^{(n-m)\Delta t/\tau}
\end{align}

\begin{figure}[t]
\includegraphics[width=0.45\textwidth]{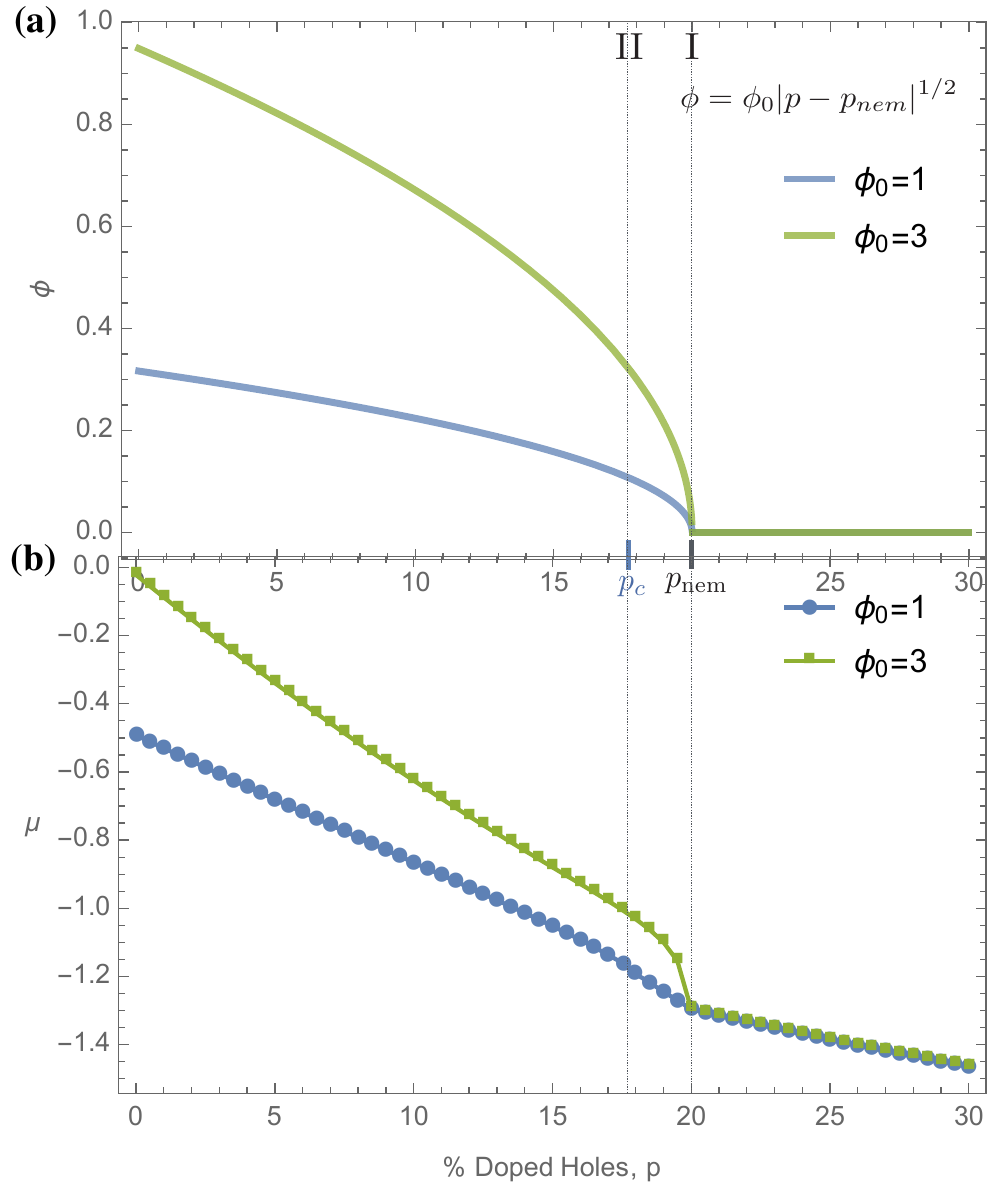}
\caption{The chemical potential as a function of doping as the nematic order parameter onsets with differing strengths $\phi_0$. While the Lifshitz transition from closed hole pockets to open sheets virtually coincides with $p_{\text{nem}} = 20\%$ for strong nematic onset (Green curves $\phi_0 = 3$), it occurs at $p_{c} < p_{\text{nem}}$ when the nematic onset is weaker (Blue curves, $\phi_0= 1$). }  \label{fig:chemVp}
\end{figure}

In Fig. 3 of the main text, we calculated the Hall number as a nematic order parameter onset at a function of hole doping $p$, $\phi(p) = \phi_0 | p - p_{nem}|^{1/2}$. To maintain the correct doping, we also (numerically) determine the chemical potential a function of $p$, as is shown in Fig.~\ref{fig:chemVp}.

\begin{figure}[t]
\includegraphics[width=0.85\textwidth]{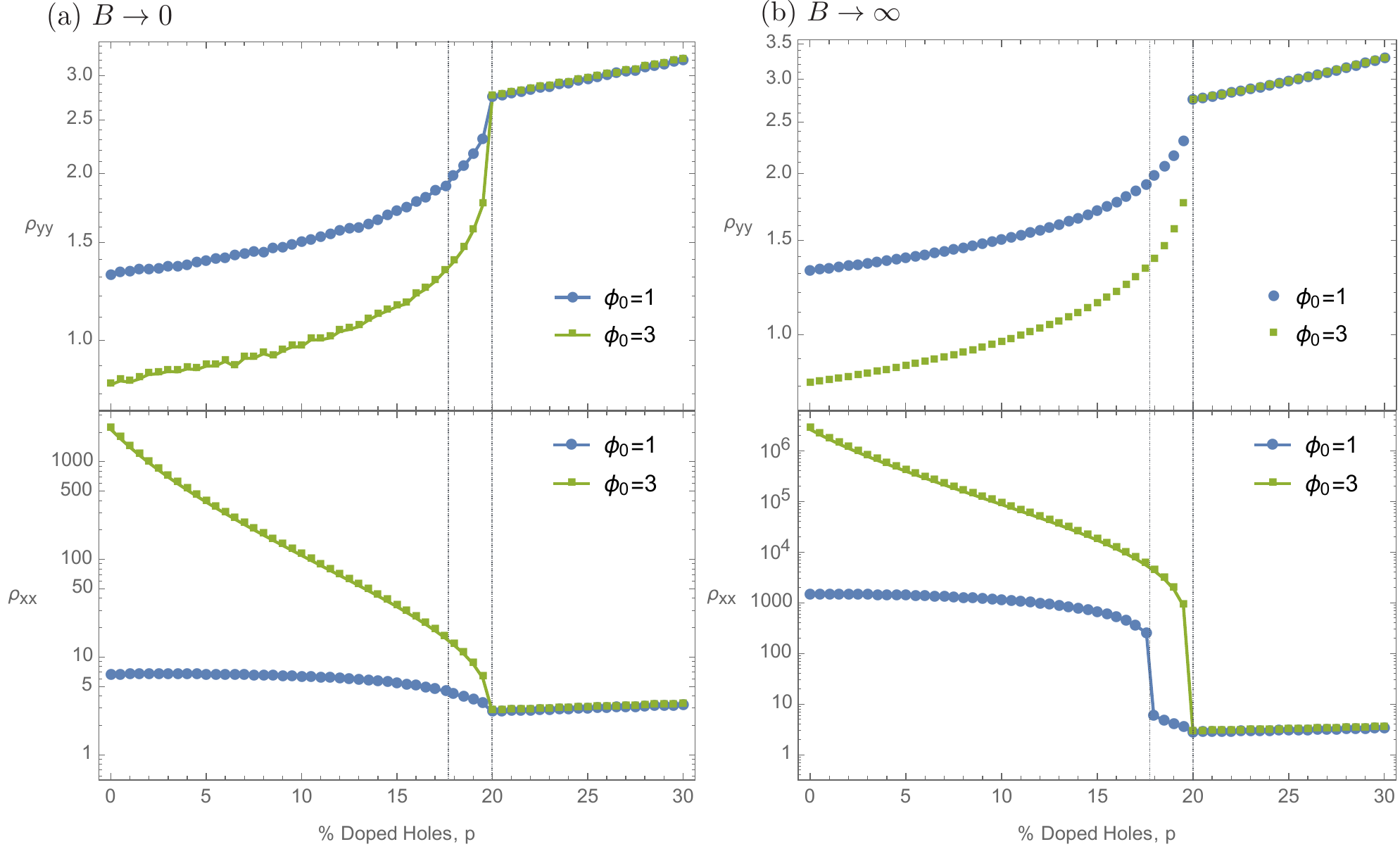}
\caption{ The longitudinal resitivities (units of $1/e^{2}\tau$, logarithmic scales) in the low (\textbf{Left}) and high field  (\textbf{Right}) limits as a function of doping.  There is a large increase in $\rho_xx$ when the Fermi surface becomes open along the $x$ direction (i.e. the nematicity forces the system to become quasi-one dimensional.\textit{ \textbf{Note:} $\rho_{xx}$ technically diverges in the infinite field limit due to the quadratic dependence, $\rho_{xx} \propto B^2$. The figure shows its value for large, but not infinite fields. }}  \label{fig:longitudinal}
\end{figure}

Meanwhile the longitudinal resistivities are shown in Fig.~\ref{fig:longitudinal}. While there is a small decrease in $\rho_{yy}$, there is a large increase in $\rho_{xx}$ in both the low and high field limits. This is natural when we realize that the quasi one-dimensional limit is being approached with increasing nematicity.



\FloatBarrier
\section{Magnetotransport in the nearest neighbor tight binding model}


\subsection{Solving the equations of motion}
To find exact expressions for the magneto-conductivity, we must solve the semiclassical equation of motion
\begin{align}
\frac{d\bm{k}}{d t} &= -\frac{e}{\hbar c} \bm{v}(\k) \times \bm{B}(\bm{r},t)
\end{align}
where $v(\k) = \partial_{\k}\varepsilon(\k)$, for a given bandstructure $\varepsilon(\k)$.
At zero temperature we are only interested in particles at the Fermi level, for the 2 dimensional nearest neighbor tight binding dispersion $\varepsilon(\k) = -2 t_x \cos{k_x} - 2 t_y \cos{k_y} - \mu$. For a $z$ directed magnetic field, $\bm{B} = B \hat{\bm{z}}$, the semiclassical equations of motion are
\begin{align}
\frac{dk_x}{dt} &= 2t_ye B \sin{k_y}\label{eq:dkx}\\
\frac{dk_y}{dt} &= -2t_x eB \sin{k_x}
\end{align}
Because the quasiparticles are always constrained to move on the Fermi surface, it is useful to eliminate $k_y$, by using the constraint that the momenta are always confined to the Fermi surface:
\begin{align}
\mu &= -2t_x \cos{k_x(t)} -2 t_y \cos{k_y(t)} \\
\implies 1 &= \left(\frac{\mu_0 - \mu_c}{\mu_0 + \mu}\right)\sin^{2}{\left[\frac{k_x(t)}{2}\right]} + \left(\frac{\mu_0 + \mu_c}{\mu_0+ \mu}\right)\sin^{2}{\left[\frac{k_y(t)}{2}\right]}
\end{align}
where $\mu_0 = 2(t_y + t_x)$ and $\mu_c = 2(t_y - t_x)$. This re-writing makes it clear that the solutions will be generalized version of ellipses, and the equation of motion for $k_x(t)$ becomes 
\begin{align}
\frac{d(k_x/2)}{du}&= \left[ 1 - \left(\frac{\mu_0 - \mu_c}{\mu_0 + \mu}\right)\sin^{2}(k_x/2)\right]^{1/2} \left[ 1 + \left(\frac{\mu_0 - \mu_c}{\mu - \mu_c}\right)\sin^{2}(k_x/2)\right]^{1/2}.
\end{align}
where $u = eB\sqrt{(\mu + 2t_x)^2 - (2t_y)^2}\,\,t$. The solutions to this non-linear equation of motion depend on the boundary conditions (see Table~\ref{table:1}), and are summarized as 
\begin{equation}
 k_x(t) =
  \begin{cases}
    2\tan^{-1}\left[\frac{m_0 v_{y0}}{2} \,\sd(\omega_0 t,\kappa)\right],    & \quad \text{for electron pockets, } \mu \le -2(t_y-t_x)\\
    \\
    2\tan^{-1}\left[\frac{m_0 v_{y0}}{2\kappa}  \,\sc(\kappa\,\omega_0\, t,1/\kappa)\right],    & \quad \text{for open Fermi surfaces, } |\mu| \le 2(t_y-t_x)\\
    \\
    \pi + 2\tan^{-1}\left[\frac{m_0 v_{y\pi}}{2}\,\sd( \omega_0 t,\kappa)\right],    & \quad \text{for hole pockets, } \mu \ge 2(t_y-t_x)\\
  \end{cases}
\end{equation}
\begin{table*}\centering
\ra{1.3}
\begin{tabular}{ l  l  l  l }
\toprule		
 & Closed Electron Pockets & Open Fermi surface & Closed Hole pockets \\
  \midrule
 Chemical potential &$ 2(t_y + t_x) \le \mu \le -2(t_y - t_x)$ & $|\mu| \le 2(t_y - t_x)$  & $-2(t_y + t_x) \le \mu \le -2(t_y - t_x) $ \\
  $k_x(t=0)$ & 0 & 0 & $\pi$\\
   $ k_y(t=0)$ &$\cos^{-1}\left(\frac{\mu + 2t_x}{-2t_y}\right)$ & $\cos^{-1}\left(\frac{\mu + 2t_x}{-2t_y}\right)$ & $\cos^{-1}\left(\frac{\mu - 2t_x}{-2t_y}\right)$  \\
   $v_x(t=0)$ & 0 & 0 & 0 \\
   $v_y(t=0)$ &$v_{y0} = \sqrt{(2t_y)^2 - (\mu + 2t_x)^2}$ & $\sqrt{(2t_y)^2 - (\mu + 2t_x)^2}$ & $v_{y\pi} = \sqrt{(2t_y)^2 - (\mu - 2t_x)^2}$ \\
  \bottomrule
  \end{tabular}
  \caption{The different types of Fermi surfaces, and initial conditions for the equations of motion in each scenario}
      \label{table:1} 
\end{table*}
Here, we have defined the `bare' cyclotron frequency $\omega_0 = eB\sqrt{4t_x t_y}$, and the elliptic modulus is (as in the main text), $\kappa = \sqrt{(\mu^2_0 - \mu^2)/(\mu^2_0 - \mu^2_c)}$. The true cyclotron frequencies are given in Eq. 6 of the main text.
\FloatBarrier
\subsection{Solutions for quasiparticle velocities}
The velocities $v_x(t)$ and $v_y(t)$ are obtained by using the equations of motion, $v_x(t) = 2t_x \sin{k_x(t)}$ and $v_y(t) = \frac{1}{eB} \frac{dk_x(t)}{dt}$. We therefore obtain 
\begin{equation}
 v_x(t) =
  \begin{cases}
    \sqrt{\frac{t_x}{t_y}} v_{y0} \frac{\sn{(\omega_0 t, \kappa)}\dn{(\omega_0 t,\kappa)}}{1+(\frac{1}{4}m^2_0 v^2_{y0} - \kappa^2)\sn^2{(\omega_0 t, \kappa)}}  & \quad \mu \le -2(t_y-t_x)\\
    \\
   \sqrt{\frac{t_x}{t_y}}  \frac{v_{y0}}{\kappa} \frac{\sn{(\kappa\,\omega_0 t, 1/\kappa)}\cn{(\kappa\,\omega_0 t,1/\kappa)}}{1+(\frac{1}{4\kappa^2}m^2_0 v^2_{y0} - 1)\sn^2{(\kappa\, \omega_0 t,1/ \kappa)}}  &\quad  |\mu| \le 2(t_y-t_x)\\
    \\
    -\sqrt{\frac{t_x}{t_y}}v_{y\pi} \frac{\sn{(\omega_0 t, \kappa)}\dn{(\omega_0 t,\kappa)}}{1+(\frac{1}{4}m^2_0 v^2_{y\pi} - \kappa^2)\sn^2{(\omega_0 t, \kappa)}}    & \quad  \mu \ge 2(t_y-t_x)\\
  \end{cases}
  \label{eq:closedv}
\end{equation}
While for the $y-$ velocities we find
\begin{equation}
 v_y(t) =
  \begin{cases}
v_{y0} \frac{\cn{(\omega_0 t, \kappa)}}{1+(\frac{1}{4}m^2_0 v^2_{y0} - \kappa^2)\sn^2{(\omega_0 t, \kappa)}}  & \quad  \mu \le -2(t_y-t_x)\\
    \\
v_{y0}\frac{\dn{(\kappa\,\omega_0 t,1/\kappa)}}{1+(\frac{1}{4\kappa^2}m^2_0 v^2_{y0} - 1)\sn^2{(\kappa\, \omega_0 t,1/ \kappa)}}  & \quad |\mu| \le 2(t_y-t_x)\\
    \\
   v_{y\pi} \frac{\cn{(\omega_0 t,\kappa)}}{1+(\frac{1}{4}m^2_0 v^2_{y\pi} - \kappa^2)\sn^2{(\omega_0 t, \kappa)}}    & \quad \mu \ge 2(t_y-t_x)\\
  \end{cases}
\label{eq:openv}
\end{equation}


\subsection{Fourier series expansions}
\label{sec:Fourier}
The solutions for the quasiparticle velocities are periodic functions of time. Thus, their Fourier series expansions are especially useful for evaluating the Chambers' integral exactly. While the Fourier series expansions for simple combinations of Jacobian elliptic functions are well known, the expansion for these rational fractions of elliptic functions are not as readily available. In Section~\ref{sec:fourierappendix} we explicitly derive these expressions by contour integration. Here, we list the results for the velocities,
\begin{align}
v^i_x(t) &= (1-2\delta_{i,h})\frac{2\pi}{m_0 K(\kappa)} \sum^{\infty}_{n=1}  \sech{\left[\frac{(2n-1)\pi K^{\prime}}{2K}\right]}\,\, \sin{\left[\frac{(2n-1)\pi u_i}{2K}\right]} \sin\left[ \frac{\left( 2n - 1\right)\pi \omega_0 t}{ 2K(\kappa)}\right]\\
v^i_y(t) &= \frac{2\pi}{m_0 K(\kappa)} \sum^{\infty}_{n=1}  \sech{\left[\frac{(2n-1)\pi K^{\prime}}{2K}\right]}\,\, \cos{\left[\frac{(2n-1)\pi u_i}{2K}\right]} \cos\left[ \frac{\left( 2n - 1\right)\pi \omega_0 t}{ 2K(\kappa)}\right]
\end{align}
for closed pockets (either $i=e$ for electron or $i=h$ for hole pockets), while for open surfaces we have
\begin{align}
v^{o}_x(t) &= \frac{2\pi \kappa}{m_0 K(1/\kappa)} \sum^{\infty}_{n=1} \sech{\left[\frac{n\pi K^{\prime}}{K}\right]}\,\, \sin{\left[\frac{n\pi u_o}{K}\right]}  \sin\left[ \frac{n \pi \kappa \omega_0 t}{K(1/\kappa)}\right]\\
v^{o}_y(t) &= \frac{2\pi \kappa}{m_0 K(1/\kappa)}\left\{\frac{1}{2} +  \sum^{\infty}_{n=1} \sech{\left[\frac{n\pi K^{\prime}}{K}\right]}\,\, \cos{\left[\frac{n\pi u_o}{K}\right]} \cos\left[ \frac{n \pi \kappa \omega_0 t}{K(1/\kappa)}\right]\right\}
\end{align}
where $\kappa$ and $\omega_c$ have their definitions as before, and $m_0 = 1/\sqrt{4t_x t_y}$, where the parameters $u_{i}$ are given by 
\begin{align}
\sn(u_e,\kappa) &=\sqrt{\frac{4t_x}{2(t_x + t_y) - \mu}} = \sqrt{\frac{\mu_0 - \mu_c}{\mu_0 - \mu} }\\
\sn(u_h,\kappa) &=\sqrt{\frac{4t_x}{2(t_x + t_y) + \mu}} = \sqrt{\frac{\mu_0 - \mu_c}{\mu_0 + \mu} }\\
\sn(\kappa u_0,1/\kappa) &=\sqrt{\frac{2(t_x + t_y) + \mu}{4t_y}} = \sqrt{\frac{\mu_0 + \mu}{\mu_0 + \mu_c} }
\end{align}


\subsection{Solutions for the conductivity}
Armed with the Fourier expansions for the conductivities we finally integrate the zero temperature Chambers' expression exactly. We first demonstrate how the finite temperature Chamber's expression can be massaged into the form given in Eq. 3 of the main text.
\begin{align}
\sigma_{\alpha\beta} &= \frac{e^{2}}{\hbar} \int \frac{d^{2}k}{(2\pi)^{2}} v_{\alpha}(\k(0))  \int^{0}_{-\infty}d\tp \left(-\frac{\partial f^{(0)}}{\partial \varepsilon}\right) v_{\beta}(\k(\tp)) e^{\tp/\tau}\\
&=_{T\rightarrow 0}\frac{e^2}{4\pi^2} \int_{FS} \frac{d\k}{|\v_F|}v_{\alpha}(\k(0))\int^{0}_{-\infty} d\tp v_{\beta}(\k(\tp))e^{\tp/\tau}\nonumber\\
&= \frac{e^2}{4\pi^2} \int^{T}_{0} \frac{dt\sqrt{\dot{k}^2_{x}+ \dot{k}^2_{y}}}{|\v_F|}v_{\alpha}(t)\int^{t}_{-\infty} d\tp v_{\beta}(t+\tp)e^{\tp/\tau}\nonumber\\
\sigma_{\alpha\beta}&= \frac{e^3B}{4\pi^2} \int^{T}_{0} dt\,\, v_{\alpha}(t)\int^{t}_{-\infty} d\tp v_{\beta}(\tp)e^{(\tp-t)/\tau}
\end{align}
Where in getting to the last line, we used the Lorentz force law. Using the Fourier series expansion for the velocities, we can do the integral over $t^{\prime}$ and then use Fourier orthogonality to perform the integral over $t$. 

We demonstrate this procedure for the longitudinal conductivity of a closed electron pocket. Schematically, writing 
\begin{align}
v_x(t) &= {\vxt}^{i} \sum^{\infty}_{n=1} a^{i}_n \sin\left[ \left( n - \frac{1}{2}\right)\frac{\pi \omega_0 t}{ K(\kappa)}\right]
\end{align}
we have for $\sigma_{xx}$:
\begin{align}
\sigma_{xx} &= \frac{e^3 B}{4\pi^2 } (\vxt^{i})^2 \int^{4K(\kappa)/\omega_0}_{0} dt \sum_{n,m} a_n a_m \sin\left[ \left( n - \frac{1}{2}\right)\frac{\pi \omega_0 t}{ K(\kappa)}\right] \int^{t}_{-\infty} dt^{\prime}\,\, \sin\left[ \left( m - \frac{1}{2}\right)\frac{\pi \omega_0 t}{ K(\kappa)}\right] e^{(\tp - t)/\tau}\nonumber\\
 &= \frac{e^2 }{4\pi^2}m_0 \omega_0 \left(\frac{2K}{\pi \omega_0}\right)^2 (\vxt^{i})^2 \int^{2\pi}_{0} du \sum_{n,m} a_n a_m \sin\left[ (2n-1)u\right] \int^{u}_{-\infty}d\up \sin\left[ (2m- 1)\up \right] e^{2K(\up - u)/\pi \omega_0 \tau}\nonumber\\
&= \frac{e^2 }{4\pi^2}m_0 \omega_0 \left(\frac{2K}{\pi \omega_0}\right)^2 (\vxt^{i})^2 \int^{2\pi}_{0} du \sum_{n,m} a_n a_m \sin\left[ (2n-1)u\right] \left(\frac{\pi\omega_0\tau}{2K}\right)\frac{\sin\left[ (2m- 1)u \right]}{1 + (2m-1)^2\left(\frac{\pi\omega_0 \tau}{2K}\right)^2}\nonumber\\
&= \frac{e^2\tau }{2\pi^2} m_0 K(\kappa)(\vxt^{i})^2    \sum^{\infty}_{n=1}  \, \,\frac{a^{2}_n}{1 + (n-\frac{1}{2})^2\left(\frac{\pi\omega_0 \tau}{K}\right)^2}\nonumber\\
&= \frac{2e^2\tau}{m_0 K(\kappa)}\sum^{\infty}_{n=1}\frac{\sech^{2}{\left(\frac{(2n-1)\pi K^{\prime}}{2K}\right)}\sin^2{\left(\frac{(2n-1)\pi u_{i}}{2K}\right)}}{1 +\left(n\omega_c \tau \right)^2},
\end{align}
where in the last line we have restored $a_n$ and $\tilde{v}^{i}_x$. Note that in going from the second to the third line above we have only kept the $\sin$ term since this is the only term which contributes upon integrating over $t$. A similar set of manipulations leads for the other elements of the conductivity tensor, and for all the different types of Fermi surfaces, leads to the formulas in Equations 8 through 10 of the main text for the conductivity tensor.


\section{High field limit of the Hall number near the Lifshitz transition}
The high field hall number is given by 
\begin{align}
\frac{1}{n_{\text{Hall}}} = \lim_{B\rightarrow \infty}  \frac{1}{B}\rho_{xy} = \lim_{B\rightarrow \infty}  \frac{1}{B}\frac{-\sigma_{xy} }{\sigma_{xx}\sigma_{yy} + \sigma^{2}_{xy}}
\end{align}
Using the expressions we derived above, for closed pockets ($i=e,h$ for electron and hole pockets respectively), we have
\begin{align}
n^{e}_{\text{Hall}} = -\frac{1}{\pi}\sum^{\infty}_{n=1} \frac{1}{\left( n - \frac{1}{2}\right)}\sech^{2}\left[\left(n - \frac{1}{2}\right)\frac{\pi K^{\prime}(\kappa)}{K(\kappa)}\right]\sin\left[\left(2n-1\right)\frac{\pi u_{e}}{K(\kappa)}\right]\\
n^{h}_{\text{Hall}} = \frac{1}{\pi}\sum^{\infty}_{n=1} \frac{1}{\left( n - \frac{1}{2}\right)}\sech^{2}\left[\left(n - \frac{1}{2}\right)\frac{\pi K^{\prime}(\kappa)}{K(\kappa)}\right]\sin\left[\left(2n-1\right)\frac{\pi u_{h}}{K(\kappa)}\right]
\end{align}
while for open pockets $i=o$, the expression is 
\begin{align}
n^{o}_{\text{Hall}} = -\frac{\frac{1}{\pi^2}\sum^{\infty}_{n=1} \frac{1}{n^2}\sech^{2}\left[\frac{n\pi K^{\prime}(1/\kappa)}{K(1/\kappa)}\right]\sin^{2}\left[\frac{n\pi u_{o}}{K(1/\kappa)}\right]}{\frac{1}{\pi}\sum^{\infty}_{n=1} \frac{1}{2n}\sech^{2}\left[\frac{n\pi K^{\prime}(1/\kappa)}{K(1/\kappa)}\right]\sin\left[\frac{2n\pi u_{o}}{K(1/\kappa)}\right]} - \frac{1}{\pi}\sum^{\infty}_{n=1} \frac{1}{n}\sech^{2}\left[\frac{n\pi K^{\prime}(1/\kappa)}{K(1/\kappa)}\right]\sin\left[\frac{2n\pi u_{o}}{K(1/\kappa)}\right]\label{eq:openinf}
\end{align}.


\subsection{Closed pockets}
To make progress note that each of the infinite sums looks like a Fourier series expansion. In fact, the coefficient $\sech[(n-1/2)\pi K^{\prime}/K]$ appears in the Fourier series expansion for $\cn(u,k)$:
\begin{align}
\cn(u,k) &= \frac{\pi}{K k} \sum^{\infty}_{n=1} \sech\left[\left(n-\frac{1}{2}\right)\frac{\pi K^{\prime}}{K}\right] \cos\left[\left(n - \frac{1}{2}\right) \frac{\pi u}{K}\right]
\end{align}
A convolution of two Jacobian $\cn$ functions, followed by two integrals allows us to re-express these infinite sums as analytic expressions, albeit involving integrals that cannot be performed. The results for the closed (electron and hole) pockets are
\begin{align}
 n^{e/h}_{\text{Hall}} &= \mp \frac{1}{\pi}\sum^{\infty}_{n=1} \frac{1}{\left( n - \frac{1}{2}\right)}\sech^{2}\left[\left(n - \frac{1}{2}\right)\frac{\pi K^{\prime}(\kappa)}{K(\kappa)}\right]\sin\left[\left(2n-1\right)\frac{\pi u_{e/h}}{K(\kappa)}\right] \\
&= \mp\frac{2\kappa}{\pi^2} \int^{\pi/2}_{0}d\theta\,\,\frac{\cos{\theta}}{\sqrt{1-\kappa^2\sin^{2}\theta}}\,\tan^{-1}{\left( \frac{\kappa\, \sn(2u_{e/h},\kappa)}{\dn(2u_{e/h},\kappa)} \cos{\theta}\right)}\\
&= \mp\frac{1}{2\pi^2}\sqrt{\frac{(2t_x + 2t_y)^2 - \mu^2}{t_x t_y}}\int^{\pi/2}_{0} d\theta\,\,\frac{\cos{\theta}}{\sqrt{1-\kappa^2\sin^{2}\theta}}\,\tan^{-1}{\left( \mp\sqrt{\frac{(2t_x + 2t_y)^2 - \mu^2}{\mu^2}}  \cos{\theta}\right)}\\
\end{align}
This is in fact exactly the density of the metal (modulo 2), as can be demonstrated by taking  the derivative w.r.t. $\mu$, to yield the density of states. We have
\begin{align}
\rho(\mu) = \frac{dn_{e/h}}{d\mu} = \frac{1}{2\pi^2\sqrt{t_xt_y}} K\left(\sqrt{\frac{(2t_x + 2t_y)^2 - \mu^2}{16t_x t_y}}\right)  = \frac{2}{\pi^2\sqrt{\mu^2_0 - \mu^2_c}}K\left(\sqrt{\frac{\mu^2_0 - \mu^2}{\mu^2_0 - \mu^2_c}}\right)
\end{align}
which is the well known expression for the density of states of a 2d tight binding model. For $\mu_0 = \mu_c + \delta\mu $ this diverges logarithmically like:
\begin{align}
\rho(\mu = \mu_c + \delta\mu) = -\frac{1}{\pi^2\sqrt{\mu^2_0 - \mu^2_c}}\log{\left[\frac{ \mu_c \delta\mu}{8(\mu^2_0 - \mu^2_c)}\right]}
\end{align}


\subsection{Open sheets}
The expression (Eq.~\ref{eq:openinf}) for the hall number of an open Fermi surface at infinite field involves two related sums:
\begin{align}
s_1(u_o) &= \frac{1}{\pi}\sum^{\infty}_{n=1} \frac{1}{n}\sech^{2}\left[\frac{n\pi K^{\prime}(1/\kappa)}{K(1/\kappa)}\right]\sin\left[\frac{2n\pi u_{o}}{K(1/\kappa)}\right]\\
s_2(u_o) &= \frac{1}{\pi^2}\sum^{\infty}_{n=1} \frac{1}{n^2}\sech^{2}\left[\frac{n\pi K^{\prime}(1/\kappa)}{K(1/\kappa)}\right]\sin^{2}\left[\frac{n\pi u_{o}}{K(1/\kappa)}\right]
\end{align}
It is clear that
\begin{align}
s_2(u_o) = \frac{1}{K(1/\kappa)} \int^{u_o}_{0} du\,\, s_1(u)
\end{align}
Once more, use the Fourier expansion of an elliptic function:
\begin{align}
\dn(u,k) &= \frac{\pi}{2K} + \frac{\pi}{K} \sum^{\infty}_{n=1}\sech\left[\frac{n\pi K^{\prime}}{K}\right]\sin\left[\frac{n\pi u_{o}}{K}\right]
\end{align}
along with a convolution followed by an integral w.r.t. $u$, to give:
\begin{align}
s_1(u_o) &=  \frac{1}{\pi}\sum^{\infty}_{n=1} \frac{1}{n}\sech^{2}\left[\frac{n\pi K^{\prime}(1/\kappa)}{K(1/\kappa)}\right]\sin\left[\frac{2n\pi u_{o}}{K(1/\kappa)}\right]\\
&= 1 - \frac{u_o}{K(1/\kappa)} + \frac{2}{\pi^2}\int^{\pi/2}_{0}d\theta\,\,\, \tan^{-1}\left[\frac{\sn(2u_0,1/\kappa)}{\cn(2u_0,1/\kappa)}\sqrt{1 - \frac{1}{\kappa^2}\sin^{2}\theta}\right] \label{eq:s1eq}
\end{align}
where the integral cannot be done in terms of elementary functions. A further integral gives 
\begin{align}
s_2(u_o) &=  \frac{1}{\pi^2}\sum^{\infty}_{n=1} \frac{1}{n^2}\sech^{2}\left[\frac{n\pi K^{\prime}(1/\kappa)}{K(1/\kappa)}\right]\sin^{2}\left[\frac{n\pi u_{o}}{K(1/\kappa)}\right]\\
&= \frac{u_o}{K(1/\kappa)} - \frac{u^{2}_0}{2K^{2}(1/\kappa)} + \frac{2}{\pi^2 K(1/\kappa)}\int^{u_o}_0 du \int^{\pi/2}_{0}d\theta\,\,\, \tan^{-1}\left[\frac{\sn(2u,1/\kappa)}{\cn(2u,1/\kappa)}\sqrt{1 - \frac{1}{\kappa^2}\sin^{2}\theta}\right]\label{eq:s2eq2}
\end{align}
where once more the integral cannot be expressed in terms of elementary functions.


\subsection{Asymptotic scaling at the critical point}
Despite the fact that Equations ~\ref{eq:s1eq} and ~\ref{eq:s2eq2} contain integrals which cannot be performed, the asymptotic behavior of these sums is determined by the preceding terms. Concentrating first on $s_1(u_o)$ in the limit $\mu = \mu_c - \delta\mu$, where $\kappa \rightarrow 1$, it can be shown that
\begin{align}
u_o(\mu = \mu_c -\delta\mu) &= \sn^{-1}\left[\sqrt{\frac{\mu_0 + \mu_c -\delta\mu}{\mu_0 + \mu_c}}\,\,,\,\, \sqrt{\frac{\mu^2_0 - \mu^2_c }{\mu^2_0 - (\mu_c-\delta\mu)^2}}\right] \\
&= -\tanh^{-1}\left(\sqrt{\frac{\mu_0 +\mu_c}{\mu_0 - \mu_c}}\right) - \frac{1}{2}\log{\left(\frac{-\mu_c \delta\mu}{8(\mu^2_0 - \mu^2_c)}\right)} + \O\left(\frac{\delta\mu}{\mu_c}\right)
\end{align}
This, together with the expansion for the elliptic function near to the van Hove, yields the following asymptotic expression for the ratio $u_o/K$:
\begin{align}
\frac{u_o}{K(1/\kappa)}   = 1 - \frac{\log\left[\frac{\mu_0 - \sqrt{\mu^2_0 - \mu^2_c}}{\mu_c} \right]}{\log{\left[\frac{ \mu_c \delta\mu}{8(\mu^2_0 - \mu^2_c)}\right]}} + \O\left(\frac{\delta\mu}{\mu_c}\right) \label{eq:u/k}
\end{align}
Furthermore, we find that the integral in Equation \ref{eq:s1eq} is roughly a constant in this limit, and so we can set $\kappa = 1$ and perform the integral, to yield:
\begin{align}
&\lim_{\mu\rightarrow \mu_c} \frac{2}{\pi^2}\int^{\pi/2}_{0}d\theta\,\,\, \tan^{-1}\left[\frac{\sn(2u_0,1/\kappa)}{\cn(2u_0,1/\kappa)}\sqrt{1 - \frac{1}{\kappa^2}\sin^{2}\theta}\right] \nonumber\\
&=  -\frac{2}{\pi^2} \int^{\pi/2}_0 d\theta  \tan^{-1}\left[\sqrt{\frac{\mu^2_0}{\mu^2_c}-1}(\cos{\theta})\right]\\
&= - \frac{2}{\pi^2}\left[ \frac{3}{2} \zeta(2) + \tanh^{-1}\left( \sqrt{\frac{\mu_0 -\mu_c}{\mu_0 + \mu_c}}\right)\log\left(\frac{\mu_0 -\mu_c}{\mu_0 + \mu_c}\right) + \text{Li}_{2}\left( -\frac{\mu_c}{\mu_0 + \sqrt{\mu^2_0 - \mu^2_c}}\right) -  \text{Li}_{2}\left(\frac{\mu_c}{\mu_0 + \sqrt{\mu^2_0 - \mu^2_c}}\right)\right]\\
&= - n_c(\mu_0,\mu_c)
\end{align}
where we have define $n_c$, the density at the critical point, $\zeta(2) = \pi^2/6$ is the Riemann zeta function, and $\text{Li}_n(x)$ is the polylogarithm function.

\begin{figure}
\subfigure[{ $s_1(\mu)$ with $t_x = 0.5 t_y$}]{
\includegraphics[width=0.49\textwidth]{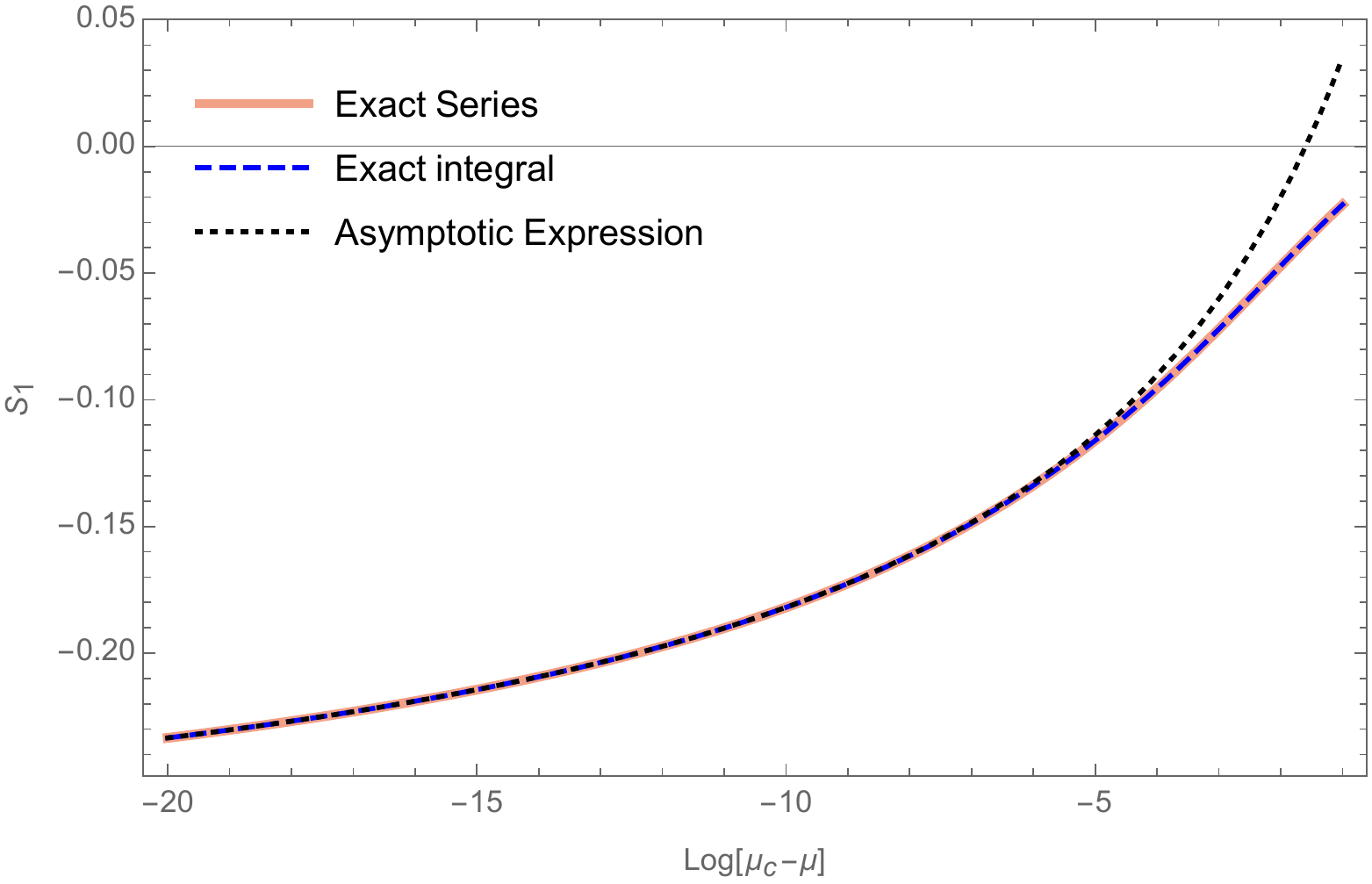}\label{fig:s1approx1}}
\subfigure[{$s_2(\mu)$ with $t_x = 0.5t_y$}]{
\includegraphics[width=0.49\textwidth]{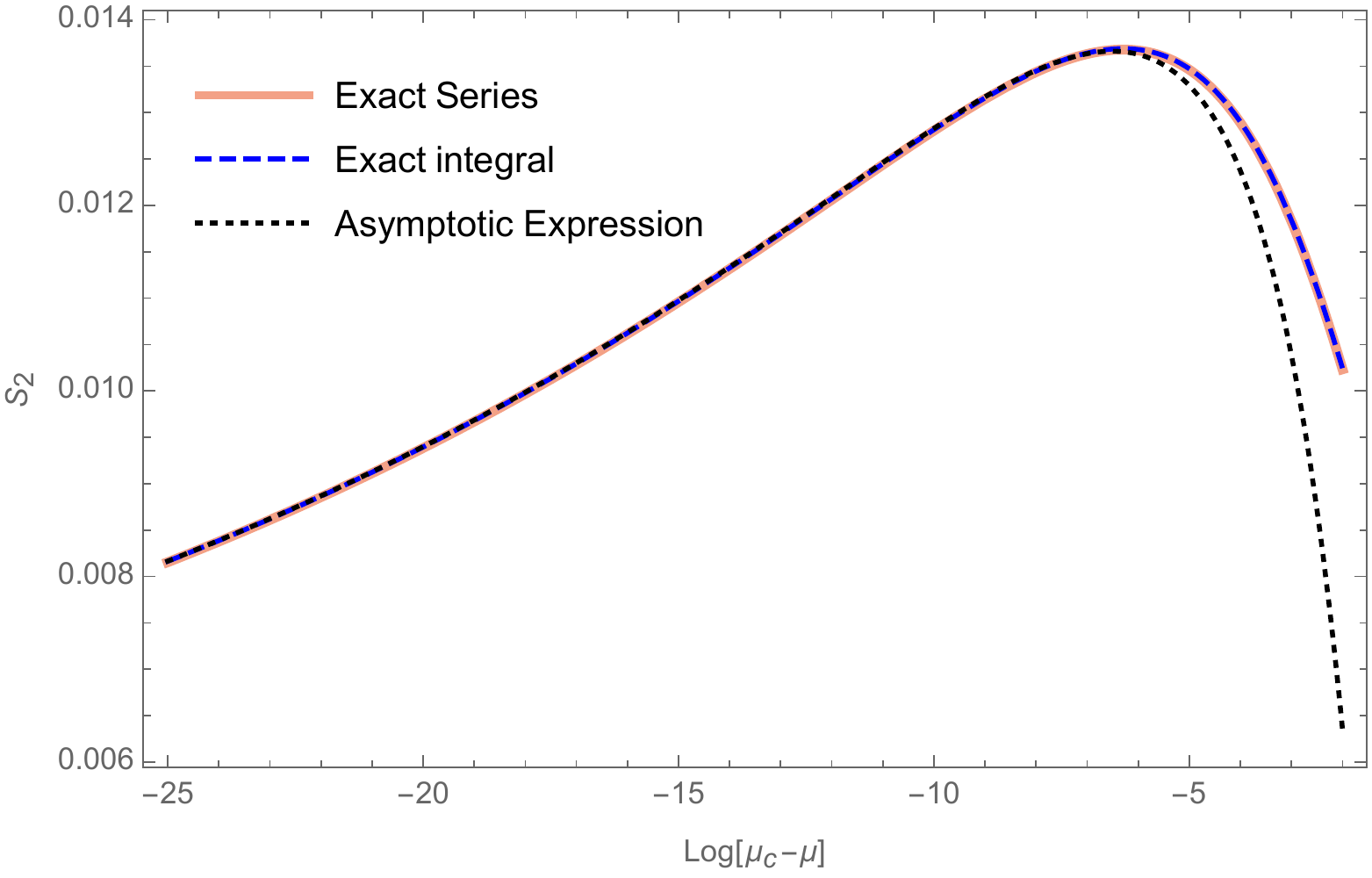}\label{fig:s2approx1}} 
\caption{The two sums which occur in the high field expression for the hall number, shown logarithmically near to the van Hove singularity.}
\end{figure}

Putting this all together, we find that this first sum in the limit $\mu \rightarrow \mu_c -\delta\mu$ is 
\begin{equation}
\lim_{\mu \rightarrow \mu_c} s_1(u_o) = - n_c(\mu_0,\mu_c) +  \frac{\log\left[\frac{\mu_0 - \sqrt{\mu^2_0 - \mu^2_c}}{\mu_c} \right]}{\log{\left[\frac{ \mu_c \delta\mu}{8(\mu^2_0 - \mu^2_c)}\right]}}
 + \O\left(\frac{\delta\mu}{\mu_c}\right)
\end{equation}
In Figure ~\ref{fig:s1approx1} we plot an exact evaluation of the sum, the exact integral representation of the sum, and the asymptotic approximation to this sum, in the limit $\mu \rightarrow \mu_c$.

\textbf{The second sum, $s_2(u)$} has $u$ dependence which is less obvious (the integral over $u$ in Equation \ref{eq:s2eq} is not a constant in the limit $\mu \rightarrow \mu_c$). Nevertheless, an analytic approximation is possible, up to a $\mu$ independent constant. We first note that the following infinite sum can be done:
\begin{align}
\sum^{\infty}_{n=1} \frac{1}{k} \sin{(k\theta)} = \frac{1}{2}\left(\pi - \theta\right)
\end{align} 
Integrating this w.r.t. $\theta$ gives the next infinite sum:
\begin{align}
\sum^{\infty}_{n=1} \frac{1}{k^2} \sin^{2}{\left(\frac{k\theta}{2}\right)} = \frac{\theta}{8}\left(2\pi - \theta\right)
\end{align} 
Using this equation, we find
\begin{align}
s_2(u_o) &=  \frac{1}{\pi^2}\sum^{\infty}_{n=1} \frac{1}{n^2}\sech^{2}\left[\frac{n\pi K^{\prime}(1/\kappa)}{K(1/\kappa)}\right]\sin^{2}\left[\frac{n\pi u_{o}}{K(1/\kappa)}\right]\\
&=  \frac{1}{\pi^2}\sum^{\infty}_{n=1} \frac{1}{n^2}\sin^{2}\left[\frac{n\pi u_{o}}{K(1/\kappa)}\right] -  \frac{1}{\pi^2}\sum^{\infty}_{n=1} \frac{1}{n^2}\tanh^{2}\left(\frac{n\pi K^{\prime}(1/\kappa)}{K(1/\kappa)}\right)\sin^{2}\left[\frac{n\pi u_{o}}{K(1/\kappa)}\right]\\
&= \frac{u_o}{2K(1/\kappa)}\left(1-\frac{u_o}{2K(1/\kappa)}\right)  -  \frac{1}{\pi^2}\sum^{\infty}_{n=1} \frac{1}{n^2}\tanh^{2}\left(\frac{n\pi K^{\prime}(1/\kappa)}{K(1/\kappa)}\right)\sin^{2}\left[\frac{n\pi u_{o}}{K(1/\kappa)}\right]\label{eq:s2eq}
\end{align}
Empirically, we find that the second term (the sum) has the form
\begin{align}
\lim_{\mu\rightarrow \mu_c}\frac{1}{\pi^2}\sum^{\infty}_{n=1} \frac{1}{n^2}\tanh^{2}\left(\frac{n\pi K^{\prime}(1/\kappa)}{K(1/\kappa)}\right)\sin^{2}\left[\frac{n\pi u_{o}}{K(1/\kappa)}\right] \approx \frac{c(\mu_0,\mu_c)}{K(1/\kappa)} + 
\end{align}
where $c(\mu_0,\mu_c)$ is a constant. Thus, using the previously obtained expansion for $u_o/K(1/\kappa)$ (Equation ~\ref{eq:u/k}), we find 
\begin{equation}
\lim_{\mu \rightarrow \mu_c} s_2(u_o)\approx  \frac{1}{2}\frac{\log\left[\frac{\mu_0 - \sqrt{\mu^2_0 - \mu^2_c}}{\mu_c} \right]}{\log{\left[\frac{ \mu_c \delta\mu}{8(\mu^2_0 - \mu^2_c)}\right]}}
- \frac{1}{2}\frac{\log^{2}\left[\frac{\mu_0 - \sqrt{\mu^2_0 - \mu^2_c}}{\mu_c} \right]}{\log^{2}{\left[\frac{ \mu_c \delta\mu}{8(\mu^2_0 - \mu^2_c)}\right]}} + \frac{2c(\mu_0,\mu_c)}{\log{\left[\frac{ \mu_c \delta\mu}{8(\mu^2_0 - \mu^2_c)}\right]}} + \O\left(\frac{\delta\mu}{\mu_c}\right)\label{eq:s2aymptotic}
\end{equation}
In Figure~S\ref{fig:s2approx1}, we show the exact series, along with the exact rewriting (Equation \ref{eq:s2eq}), and finally the asymptotic expression (Equation ~\ref{eq:s2aymptotic}).

\begin{figure}
\subfigure[{ Logarithmic scale}]{
\includegraphics[width=0.49\textwidth]{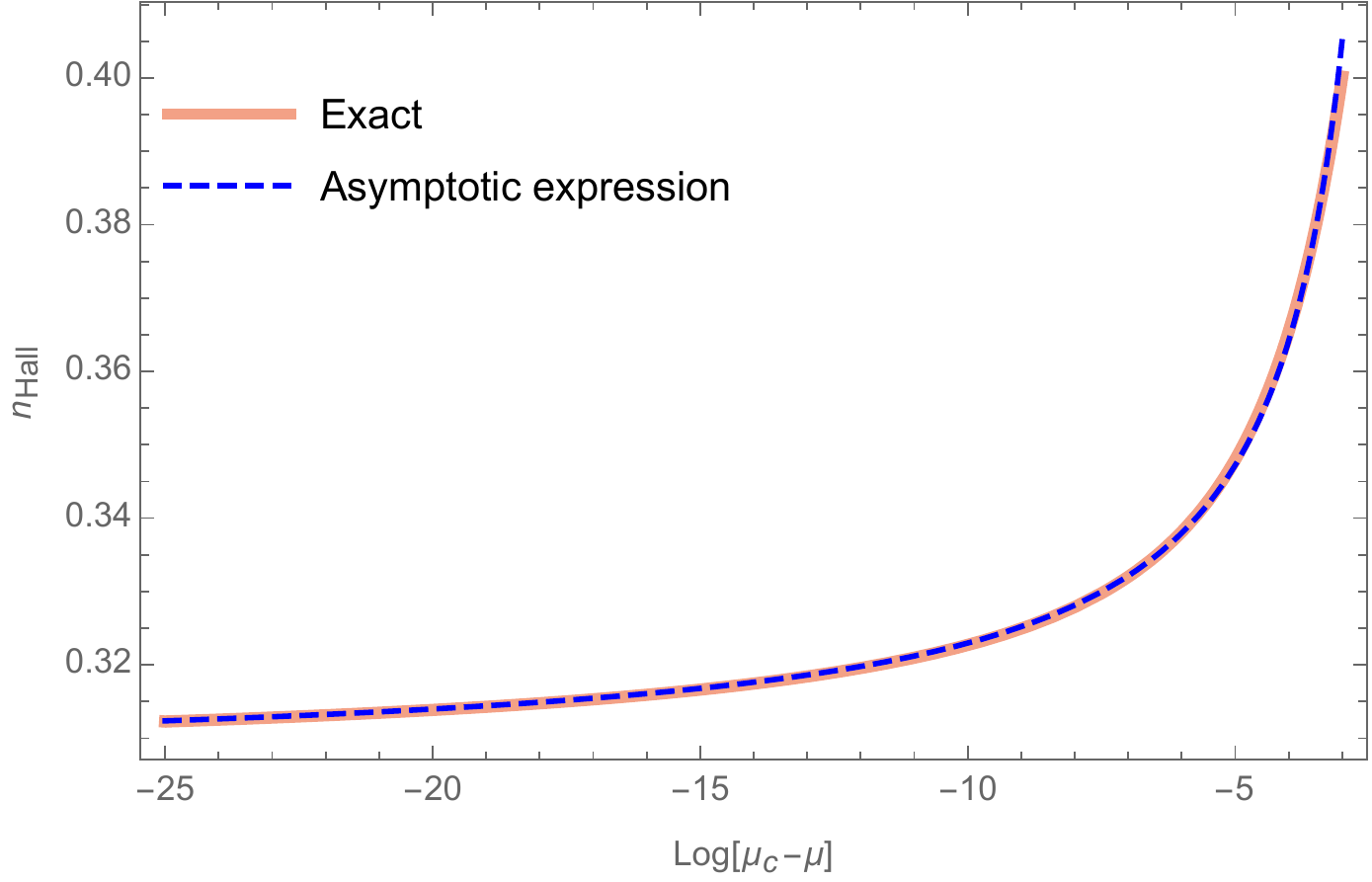}\label{fig:asmp1}}
\subfigure[{Linear scale}]{
\includegraphics[width=0.49\textwidth]{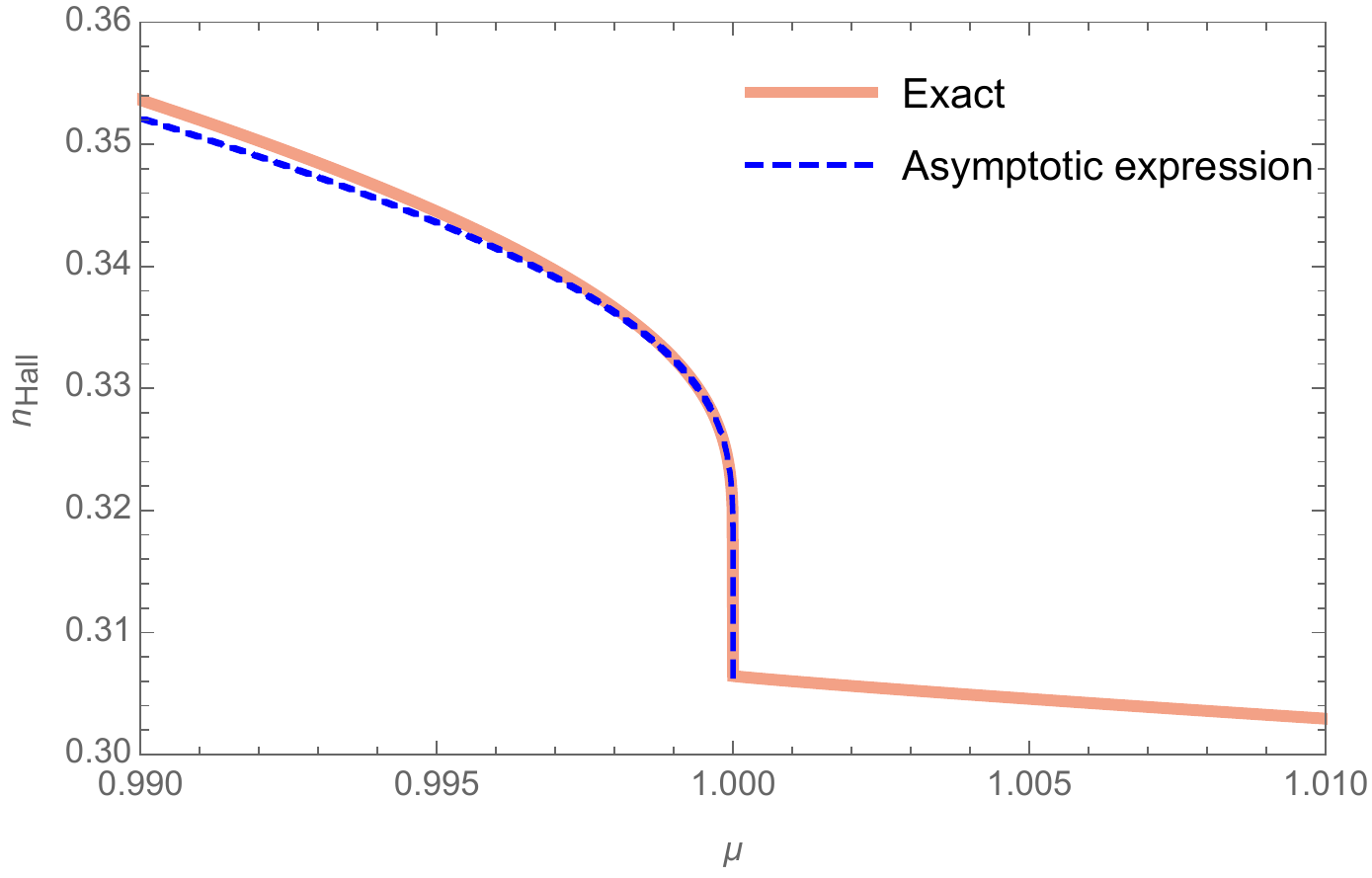}\label{fig:asymp2}} 
\caption{The asymptotic expression for the Hall number, on logarithmic and linear scales. We have chosen $t_x = 0.5$ and $t_y = 1$, so that the critical point occurs at $\mu_c = 1$. (The constant $c(\mu_0,\mu_c)$ is $ 0.2951$ for these parameters.}
\label{fig:fullasymptotics}
\end{figure}

\textbf{Putting all the results together}, we find that the Hall number in the limit $\mu \rightarrow \mu_c$ is given by
\begin{align}
\lim_{\mu \rightarrow \mu_c} n_{\text{Hall}} = \lim_{\mu \rightarrow \mu_c}\left( -\frac{s_2(u_o)}{\frac{1}{2} s_1(u_o)} - s_1(u_o)\right)
\end{align}
i.e.
\begin{equation}
\boxed{
n_{\text{Hall}}(\mu = \mu_c - \delta\mu) = n_c + \frac{(1-n_c)\log\left[\frac{\mu_0 - \sqrt{\mu^2_0 - \mu^2_c}}{\mu_c} \right] + 4 c(\mu_0, \mu_c)}{n_c\log{\left[\frac{ \mu_c \delta\mu}{8(\mu^2_0 - \mu^2_c)}\right]} - \log\left[\frac{\mu_0 - \sqrt{\mu^2_0 - \mu^2_c}}{\mu_c} \right]} + \O\left(\frac{\delta\mu}{\mu_c}\right)
}
\end{equation}
where, to recap all the terms in this expression, $\mu_0 = 2(t_y + t_x)$ is half the bandwidth, $\mu_c = 2(t_y - t_x)$ is the value of the chemical potential at which the van Hove occurs, $c(\mu_0, \mu_c)$ is constant that depends only on $t_y$ and $t_x$, and $n_c$ is the filling at the van Hove point, given by 
\begin{align}
n_c = \frac{2}{\pi^2}\left[ \frac{3}{2} \zeta(2) + \tanh^{-1}\left( \sqrt{\frac{\mu_0 -\mu_c}{\mu_0 + \mu_c}}\right)\log\left(\frac{\mu_0 -\mu_c}{\mu_0 + \mu_c}\right) + \text{Li}_{2}\left( -\frac{\mu_c}{\mu_0 + \sqrt{\mu^2_0 - \mu^2_c}}\right) -  \text{Li}_{2}\left(\frac{\mu_c}{\mu_0 + \sqrt{\mu^2_0 - \mu^2_c}}\right)\right]
\end{align}
Figure~\ref{fig:fullasymptotics} shows this function (dashed blue), compared to the asymptotic behavior of the exact expression for the hall number, on logarithmic and linear scales. 

The expression in the main text, Eq. 14, 
\begin{align}
n_{H} (\mu) - n_c = \frac{n_c\, C_1}{\log{\left|C_2\mu_c/\delta\mu\right|}} + \O\left(\frac{\delta\mu}{\mu_c}\right),
\end{align}
is obtained by setting 
\begin{align}
C_1 &= \frac{1}{n^2_c}\left[(1-n_c)\log\left[\frac{\mu_0 - \sqrt{\mu^2_0 - \mu^2_c}}{\mu_c} \right] + 4 c(\mu_0, \mu_c)\right]\\
C_2 &= \frac{\mu_c^3}{8(\mu^2_0 - \mu^2_c)(\mu_0 - \sqrt{\mu^2_0 - \mu^2_c})}
\end{align}

\section{Low field hall number near the Lifshitz transition}
In the limit of the field approaching zero, the hall number for closed Fermi surfaces is given by
\begin{align}
n^{e/h}_{\text{Hall}} = -\frac{2}{\pi}\frac{\left(\sum^{\infty}_{n = 1} \sech^{2}\left[(2n-1)\frac{\pi K^{\prime}}{2K}\right]\sin^{2}\left[(2n-1)\frac{\pi u_{e/h}}{2K}\right]\right)\left(\sum^{\infty}_{n = 1} \sech^{2}\left[(2n-1)\frac{\pi K^{\prime}}{2K}\right]\cos^{2}\left[(2n-1)\frac{\pi u_{e/h}}{2K}\right]\right)}{\sum^{\infty}_{n = 1}\left(n-\frac{1}{2}\right) \sech^{2}\left[(2n-1)\frac{\pi K^{\prime}}{2K}\right]\sin\left[(2n-1)\frac{\pi u_{e/h}}{2K}\right]\cos\left[(2n-1)\frac{\pi u_{e/h}}{2K}\right]}
\end{align}
where $K = K(\kappa)$ is the complete elliptic integral with modular parameter $\kappa$, etc., while for open Fermi surfaces the hall number is 
\begin{align}
n^{o}_{\text{Hall}} = -\frac{2}{\pi} \frac{\left(\sum^{\infty}_{n = 1} \sech^{2}\left[\frac{n\pi K^{\prime}}{K}\right]\sin^{2}\left[\frac{n\pi u_{o}}{K}\right]\right)\left(\frac{1}{2} + \sum^{\infty}_{n = 1} \sech^{2}\left[\frac{n\pi K^{\prime}}{K}\right]\cos^{2}\left[\frac{n\pi u_{o}}{K}\right]\right)}{\sum^{\infty}_{n = 1}n\,\,\sech^{2}\left[\frac{n\pi K^{\prime}}{K}\right]\sin\left[\frac{n\pi u_{o}}{K}\right]\cos\left[\frac{n\pi u_{o}}{K}\right]}
\end{align}
where now the modular parameter is $1/\kappa$. Using the same tricks/techniques of the previous section, we can in fact find exact expressions for these infinite sums, in terms of elementary functions.

Focusing on the open Fermi surface side, let us define an elementary sum
\begin{align}
s(u_o) &= \sum^{\infty}_{n=1}  \sech^{2}\left[\frac{n\pi K^{\prime}(1/\kappa)}{K(1/\kappa)}\right]\cos\left[\frac{n\pi u_{o}}{K(1/\kappa)}\right]\nonumber\\
&=-\frac{1}{2} + \frac{2K(1/\kappa)}{\pi^2} \frac{ \dn(u_o,1/\kappa)}{\sn^{2}(u_o,1/\kappa)}\left[ K(1/\kappa) - \Pi\left(\kappa^{-2}\sn^{2}(u_o,1/\kappa),1/\kappa)\cn^{2}(u_o,1/\kappa\right)\right]
\end{align}
where the second equality follows from using the convolution theorem on the Fourier series expansion of $\dn(u,k)$. Here, $\Pi$ is the complete elliptic integral of the third kind. Note that this implies 
\begin{align}
s(0) &= -\frac{1}{2} + \frac{2}{\pi^2} E(1/\kappa)K(1/\kappa)
\end{align}
where $E(k)$ is the complete elliptic integral of the second kind. With this definition, it is not difficult to see that 
\begin{align}
\sum^{\infty}_{n = 1} \sech^{2}\left[\frac{n\pi K^{\prime}}{K}\right]\sin^{2}\left[\frac{n\pi u_{o}}{K}\right] &= \frac{1}{2}\left[ s(0) - s(2u_o)\right]\\
\sum^{\infty}_{n = 1} \sech^{2}\left[\frac{n\pi K^{\prime}}{K}\right]\cos^{2}\left[\frac{n\pi u_{o}}{K}\right] &= \frac{1}{2}\left[ s(0) + s(2u_o)\right]\\
\frac{1}{2}\sum^{\infty}_{n = 1} n \,\,\sech^{2}\left[\frac{n\pi K^{\prime}}{K}\right]\sin\left[\frac{2n\pi u_{o}}{K}\right] &= -\frac{K}{2\pi} s^{\prime}(2u_o)
\end{align}
So we end up with the following expression for the Hall number in the zero field limit:
\begin{align}
n^{o}_{\text{Hall}} = \frac{1}{K(1/\kappa)}\frac{ s(0) - s(2u_o) +  s^2(0) - s^2(2u_o) }{s^{\prime}(2u_o)}
\end{align}

This expression is fairly complicated, but written in full has the form:
\begin{align}
n^{(o)}_{H} = \frac{2 \left[\left(\mu ^2-\mu _0^2\right){}^2 E\left(\frac{\mu _c^2-\mu _0^2}{\mu ^2-\mu _0^2}\right){}^2-\mu _c^2 \left(\frac{\mu ^2 \Pi \left(1-\frac{\mu _c^2}{\mu _0^2}|\frac{\mu _c^2-\mu _0^2}{\mu ^2-\mu _0^2}\right)}{\mu _0}-\mu _0 K\left(\frac{\mu _c^2-\mu _0^2}{\mu ^2-\mu _0^2}\right)\right){}^2\right]}{\pi ^2 \mu  \sqrt{\mu _0^2-\mu ^2} \left(\left(\mu _c^2-\mu ^2+\mu _0^2\right) K\left(\frac{\mu _c^2-\mu _0^2}{\mu ^2-\mu _0^2}\right)+\left(\mu ^2-\mu _0^2\right) E\left(\frac{\mu _c^2-\mu _0^2}{\mu ^2-\mu _0^2}\right)+\mu _c^2 \left(-\Pi \left(1-\frac{\mu _c^2}{\mu _0^2}|\frac{\mu _c^2-\mu _0^2}{\mu ^2-\mu _0^2}\right)\right)\right)}\end{align}
This is an exact expression for the low field Hall number. The series expansion is complicated, but in the limit $\mu\rightarrow \mu_c$, we find a weak singularity in the hall number:
\begin{align}
n^{o}_{\text{Hall}} \approx \alpha + \beta |\mu_c - \mu| \log{|\mu_c - \mu|} + \ldots
\end{align}
where $\alpha$ and $\beta$ are constants. 

\section{Fourier Series for rational fractions of Jacobian Elliptic functions}
\label{sec:fourierappendix}
We must calculate Fourier series expansions for the functions
\begin{align}
\frac{\sn(u,k)\dn(u,k)}{1- \alpha^2 \sn^{2}(u,k)},\quad \frac{\cn(u,k)}{1- \alpha^2 \sn^{2}(u,k)},\quad \frac{\sn(u,k)\cn(u,k)}{1- \alpha^2 \sn^{2}(u,k)},\,\,\,\text{and}\,\, \frac{\dn(u,k)}{1- \alpha^2 \sn^{2}(u,k)}
\end{align}
for $0 < \alpha < k < 1$. Note that the condition $\alpha < k$ follows from the forms for the quasiparticle velocities found in Equations \ref{eq:closedv} and \ref{eq:openv}. These Fourier series expansions are not readily available in the literature, so here we discuss their derivations in a little detail.

\begin{figure}
\includegraphics[width=0.55\textwidth]{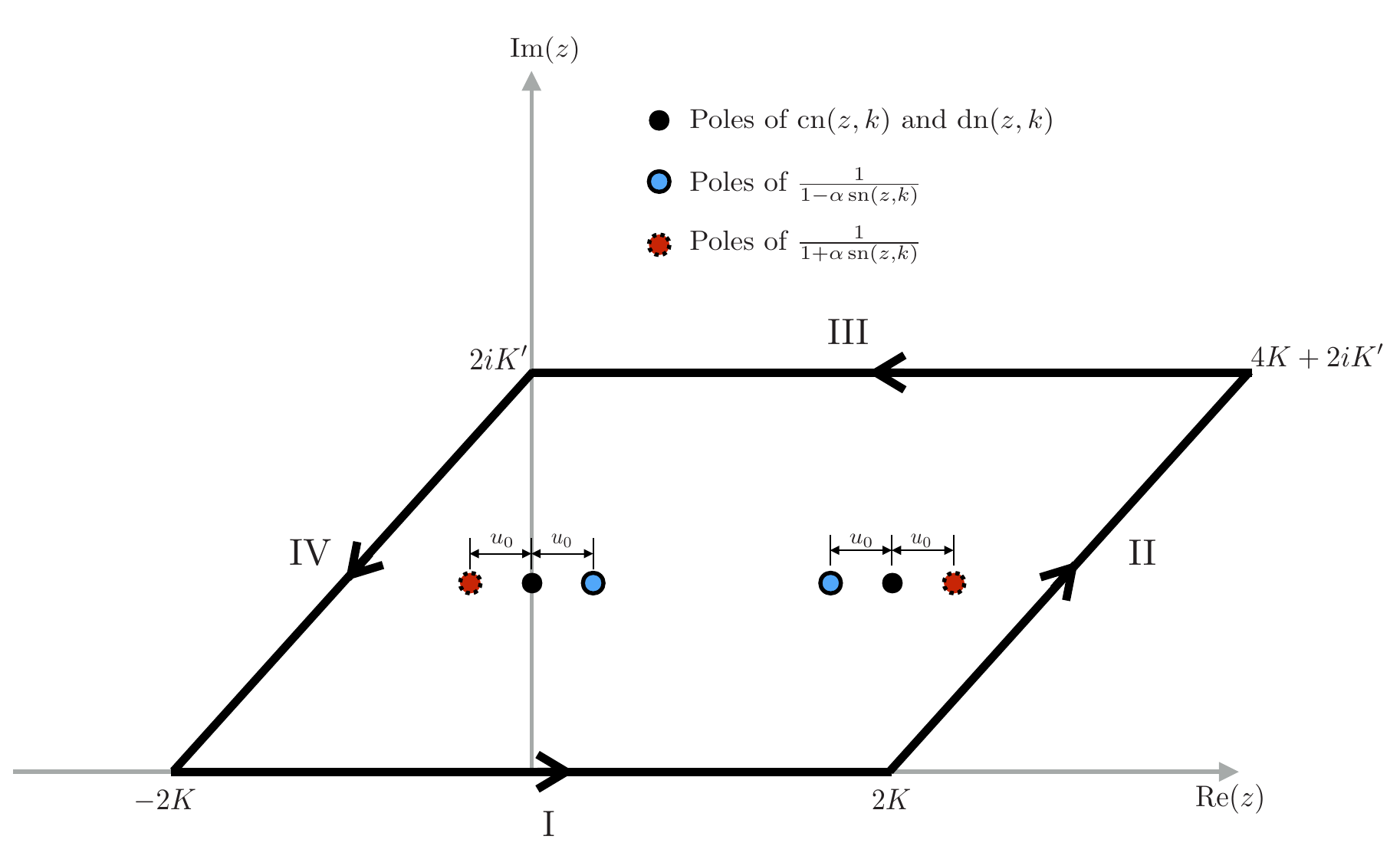}
\caption{Contour used for finding Fourier coefficients of rational fractions of Jacobian elliptic functions, with positions of poles indicated. }  \label{fig:contour1}
\end{figure}

These four Fourier series can be obtained from simple addition and subtraction of the functions 
\begin{align}
\frac{\dn(u,k)}{1 \pm \alpha\, \sn(u,k)},\,\,\text{and}\,\, \frac{\cn(u,k)}{1 \pm \alpha\, \sn(u,k)}
\end{align}
These are both periodic functions, with a period of $4K$, and we can calculate its Fourier coefficients by using the relation 
\begin{align}
a_n = c_n \int^{2K}_{-2K} e^{i\frac{n\pi u}{2K}} \frac{\text{[c/d]n}(u,k)}{1 \pm \alpha\,\sn(u,k)} du
\end{align}
with $c_n = (2iK)^{-1}$ for odd functions and $c_n = (2K)^{-1}$ for even functions. These integrals can be done by considering a contour in the complex plane, as shown in Figure ~\ref{fig:contour1}.
First let us note the positions of the poles ($\alpha >0$):
\begin{enumerate}
\item $\dn(z,k)$ has simple poles at $iK^{\prime}$ and $2K + iK^{\prime}$
\item $\cn(z,k)$ likewise has simple poles at $i\Kp$ and $2K + i\Kp$
\item $(1- \alpha\,\sn(z,k))^{-1}$ has simples poles at $u_0 + i\Kp$ and $2K - u_0 + i\Kp$
\item $(1+ \alpha\,\sn(z,k))^{-1}$ has simples poles at $-u_0 + i\Kp$ and $2K + u_0 + i\Kp$
\end{enumerate}
where $u_0$ is the solution of the equation 
\begin{align} 
\sn(u_0,k) = \frac{\alpha}{k}
\end{align}

We therefore consider a contour integral, with the contour shown in Fig.~\ref{fig:contour1}. Considering first,  $\cn(u,k)/(1\pm \alpha\,\sn(u,k))$, and defining
\begin{align}
I_{\pm\alpha} = \int^{2K}_{-2K}du \frac{\cn(u,k)e^{i\frac{n\pi u}{2K}}}{1 \pm \alpha\,\sn(u,k)} 
\end{align}
we find that $I_{\pm \alpha}$ are given by solving the simultaneous equations:
\begin{align}
I_{\alpha} + (-1)^{n+1} e^{-\frac{n\pi \Kp}{K}} I_{-\alpha} &= \frac{2\pi i}{\sqrt{\alpha^2 - k^2}} e^{-\frac{n\pi \Kp}{2K}}\left[ (-1)^{n+1}e^{i\frac{n\pi u_0}{2K}} + e^{-i\frac{n\pi u_0}{2K}}\right]\\
I_{-\alpha} + (-1)^{n+1} e^{-\frac{n\pi \Kp}{K}} I_{\alpha} &= \frac{2\pi i}{\sqrt{\alpha^2 - k^2}} e^{-\frac{n\pi \Kp}{2K}}\left[e^{i\frac{n\pi u_0}{2K}} +  (-1)^{n+1}e^{-i\frac{n\pi u_0}{2K}}\right]
\end{align}
This leads to solutions
\begin{align}
I_{\pm \alpha} = \frac{2\pi}{\sqrt{k^2 - \alpha^2}} \sech{\left(\frac{n\pi \Kp}{2K}\right)} \left[ \sin^{2}\left(\frac{n\pi}{2}\right)\cos{\left(\frac{n\pi u_0}{2K}\right)} \mp i \cos^{2}\left(\frac{n\pi}{2}\right)\sin{\left(\frac{n\pi u_0}{2K}\right)} \right]
\end{align}
So that the Fourier expansion is 
\begin{align}
\frac{\cn(u,k)}{1 \pm \alpha\, \sn(u,k)} = \frac{\pi}{K\sqrt{k^2 - \alpha^2}}\sum^{\infty}_{n=1}\sech{\left(\frac{n\pi \Kp}{2K}\right)}\left[ \sin^{2}\left(\frac{n\pi}{2}\right)\cos{\left(\frac{n\pi u_0}{2K}\right)}\cos{\left(\frac{n\pi u}{2K}\right)}\right.\nonumber\\
\quad\quad\,\,\,\,\, \left.\mp  \cos^{2}\left(\frac{n\pi}{2}\right)\sin{\left(\frac{n\pi u_0}{2K}\right)}\sin{\left(\frac{n\pi u}{2K}\right)} \right]
\label{eq:fourier1}
\end{align}

For the case of $\dn(u,k)/(1 \pm \alpha\,\sn(u,k))$, we perform a very similar computation. Once more, $\int_{II} + \int_{IV} = 0$, and defining 
\begin{align}
J_{\pm \alpha} =  \int^{2K}_{-2K}du \frac{\dn(u,k)e^{i\frac{n\pi u}{2K}}}{1 \pm \alpha\,\sn(u,k)}
\end{align}
performing the contour integrals leads to the following simultaneous equations
\begin{align}
J_{\alpha} + (-1)^{n} e^{-\frac{n\pi \Kp}{K}} J_{-\alpha} &= \frac{2\pi}{\sqrt{1-\alpha^2}} e^{-\frac{n\pi \Kp}{2K}}\left[ (-1)^{n}e^{i\frac{n\pi u_0}{2K}} + e^{-i\frac{n\pi u_0}{2K}}\right]\\
J_{-\alpha} + (-1)^{n} e^{-\frac{n\pi \Kp}{K}} J_{\alpha} &=\frac{2\pi}{\sqrt{1-\alpha^2}} e^{-\frac{n\pi \Kp}{2K}}\left[ e^{i\frac{n\pi u_0}{2K}} + (-1)^{n}e^{-i\frac{n\pi u_0}{2K}}\right]
\end{align}
whose solutions are 
\begin{align}
J_{\pm \alpha} = \frac{2\pi}{\sqrt{1 - \alpha^2}} \sech{\left(\frac{n\pi \Kp}{2K}\right)} \left[ \cos^{2}\left(\frac{n\pi}{2}\right)\cos{\left(\frac{n\pi u_0}{2K}\right)} \mp i \sin^{2}\left(\frac{n\pi}{2}\right)\sin{\left(\frac{n\pi u_0}{2K}\right)} \right]
\end{align}
We therefore find that the Fourier expansion is 
\begin{align}
\frac{\dn(u,k)}{1 \pm \alpha\, \sn(u,k)} = \frac{\pi}{K\sqrt{1 - \alpha^2}}\left\{ \frac{1}{2} + \sum^{\infty}_{n=1}\sech{\left(\frac{n\pi \Kp}{2K}\right)}\left[ \cos^{2}\left(\frac{n\pi}{2}\right)\cos{\left(\frac{n\pi u_0}{2K}\right)}\cos{\left(\frac{n\pi u}{2K}\right)} \right.\right.\nonumber\\
\quad\quad\,\,\left.\left. \mp  \sin^{2}\left(\frac{n\pi}{2}\right)\sin{\left(\frac{n\pi u_0}{2K}\right)}\sin{\left(\frac{n\pi u}{2K}\right)} \right]\right\}
\label{eq:fourier2}
\end{align}
Taking different combinations of Equations \ref{eq:fourier1} and \ref{eq:fourier2} we find our final expressions
\begin{align}
\frac{\sn(u,k)\dn(u,k)}{1 - \alpha^{2}\sn^2(u,k)}  &= \frac{\pi}{\alpha\sqrt{1-\alpha^2} K} \sum^{\infty}_{n=1} \sech{\left[\frac{(2n-1)\pi K^{\prime}}{2K}\right]}\,\, \sin{\left[\frac{(2n-1)\pi u_0}{2K}\right]} \sin{\left[\frac{(2n-1)\pi u}{2K}\right]}\\
\frac{\cn(u,k)}{1 - \alpha^{2}\sn^2(u,k)} &= \frac{\pi}{\sqrt{k^2-\alpha^2} K} \sum^{\infty}_{n=1} \sech{\left[\frac{(2n-1)\pi K^{\prime}}{2K}\right]}\,\, \cos{\left[\frac{(2n-1)\pi u_0}{2K}\right]} \cos{\left[\frac{(2n-1)\pi u}{2K}\right]}\\
\frac{\sn(u,k)\cn(u,k)}{1 - \alpha^{2}\sn^2(u,k)}  &= \frac{\pi}{\alpha \sqrt{k^2-\alpha^2} K} \sum^{\infty}_{n=1} \sech{\left[\frac{n\pi K^{\prime}}{K}\right]}\,\, \sin{\left[\frac{n\pi u_0}{K}\right]} \sin{\left[\frac{n\pi u}{K}\right]}\\
\frac{\dn(u,k)}{1 - \alpha^{2}\sn^2(u,k)} &= \frac{\pi}{\sqrt{1-\alpha^2} K} \left\{ \frac{1}{2} + \sum^{\infty}_{n=1} \sech{\left[\frac{n\pi K^{\prime}}{K}\right]}\,\, \cos{\left[\frac{n\pi u_0}{K}\right]} \cos{\left[\frac{n\pi u}{K}\right]} \right\}
\end{align}


\end{document}